\documentclass[12pt]{article}
\usepackage[dvips]{graphicx}
\usepackage{epsfig}
\usepackage{latexsym,amsmath,amsfonts,amssymb}
\usepackage{mathtools} 
\usepackage{slashed} 
\usepackage[all]{xy} 
\usepackage[makeroom]{cancel}
\usepackage[latin1]{inputenc}
\usepackage[american]{babel}
\usepackage[dvips]{graphicx}
\usepackage{bbm}
\usepackage{color}
\usepackage[unicode]{hyperref}
\def\im{Invent. Math.}

\def\a{\alpha}
\def\b{\beta}
\def\c{\gamma}
\def\d{\delta}
\def\f{\phi}               
\def\vf{\varphi}  \def\tvf{\tilde{\varphi}}
\def\vp{\varphi}
\def\g{\gamma}
\def\h{\eta}
\def\j{\psi}
\def\k{\kappa}                    
\def\l{\lambda}
\def\m{\mu}
\def\n{\nu}
\def\o{\omega}  \def\w{\omega}

\def\q{\theta}  \def\th{\theta}                  
\def\r{\rho}                                     
\def\s{\sigma}                                   
\def\t{\tau}
\def\u{\upsilon}
\def\x{\xi}
\def\z{\zeta}
\def\pt{\tilde{\varphi}}
\def\tt{\tilde{\theta}}
\def\lab{\label}
\def\6{\partial}
\def\wg{\wedge}
\def\bpsi{\bar{\psi}}
\def\bt{\bar{\theta}}
\def\bvf{\bar{\varphi}}

\DeclareMathOperator{\tr}{tr}

\newcommand{\be}{\begin{equation}}
\newcommand{\ee}{\end{equation}}
\newcommand{\beq}{\begin{equation}}
\newcommand{\eeq}{\end{equation}}
\newcommand{\bea}{\begin{eqnarray}}
\newcommand{\eea}{\end{eqnarray}}
\newcommand{\nn}{\nonumber}

\newcommand{\ba}{\begin{eqnarray}}
\newcommand{\ea}{\end{eqnarray}}

\newcommand{\beqs}{\begin{eqnarray}}
\newcommand{\eeqs}{\end{eqnarray}}
\newcommand{\bal}{\begin{aligned}}
\newcommand{\eal}{\end{aligned}}

\begin{document}
\baselineskip=15.5pt
\pagestyle{plain}
\setcounter{page}{1}


\def\del{{\partial}}
\def\vev#1{\left\langle #1 \right\rangle}
\def\cn{{\cal N}}
\def\co{{\cal O}}
\def\IC{{\mathbb C}}
\def\IR{{\mathbb R}}
\def\IZ{{\mathbb Z}}
\def\RP{{\bf RP}}
\def\CP{{\bf CP}}
\def\Poincare{{Poincar\'e }}
\def\tr{{\rm tr}}
\def\tp{{\tilde \Phi}}

\def\TL{\hfil$\displaystyle{##}$}
\def\TR{$\displaystyle{{}##}$\hfil}
\def\TC{\hfil$\displaystyle{##}$\hfil}
\def\TT{\hbox{##}}
\def\HLINE{\noalign{\vskip1\jot}\hline\noalign{\vskip1\jot}}
\def\seqalign#1#2{\vcenter{\openup1\jot
   \halign{\strut #1\cr #2 \cr}}}
\def\lbldef#1#2{\expandafter\gdef\csname #1\endcsname {#2}}
\def\eqn#1#2{\lbldef{#1}{(\ref{#1})}%
\begin{equation} #2 \label{#1} \end{equation}}
\def\eqalign#1{\vcenter{\openup1\jot
     \halign{\strut\span\TL & \span\TR\cr #1 \cr
    }}}
\def\eno#1{(\ref{#1})}
\def\href#1#2{#2}
\def\half{\frac{1}{2}}

\def\ads{{\it AdS}}
\def\adsp{{\it AdS}$_{p+2}$}
\def\cft{{\it CFT}}

\newcommand{\ber}{\begin{eqnarray}}
\newcommand{\eer}{\end{eqnarray}}

\newcommand{\beqar}{\begin{eqnarray}}
\newcommand{\cN}{{\cal N}}
\newcommand{\cO}{{\cal O}}
\newcommand{\cA}{{\cal A}}
\newcommand{\cT}{{\cal T}}
\newcommand{\cF}{{\cal F}}
\newcommand{\cC}{{\cal C}}
\newcommand{\cR}{{\cal R}}
\newcommand{\cW}{{\cal W}}
\newcommand{\eeqar}{\end{eqnarray}}
\newcommand{\tht}{\thteta}
\newcommand{\lm}{\lambda}\newcommand{\Lm}{\Lambda}
\newcommand{\R}{\mathbb{R}}


\newcommand{\nonu}{\nonumber}
\newcommand{\oh}{\displaystyle{\frac{1}{2}}}
\newcommand{\dsl}
   {\kern.06em\hbox{\raise.15ex\hbox{$/$}\kern-.56em\hbox{$\partial$}}}
\newcommand{\id}{i\!\!\not\!\partial}
\newcommand{\as}{\not\!\! A}
\newcommand{\ps}{\not\! p}
\newcommand{\ks}{\not\! k}
\newcommand{\D}{{\cal{D}}}
\newcommand{\dv}{d^2x}
\newcommand{\Z}{{\cal Z}}
\newcommand{\N}{{\cal N}}
\newcommand{\Dsl}{\not\!\! D}
\newcommand{\Bsl}{\not\!\! B}
\newcommand{\Psl}{\not\!\! P}
\newcommand{\eeqarr}{\end{eqnarray}}
\newcommand{\ZZ}{{\rm \kern 0.275em Z \kern -0.92em Z}\;}


\def\del{{\delta^{\hbox{\sevenrm B}}}} \def\ex{{\hbox{\rm e}}}
\def\azb{A_{\bar z}} \def\az{A_z} \def\bzb{B_{\bar z}} \def\bz{B_z}
\def\czb{C_{\bar z}} \def\cz{C_z} \def\dzb{D_{\bar z}} \def\dz{D_z}
\def\im{{\hbox{\rm Im}}} \def\mod{{\hbox{\rm mod}}} \def\tr{{\hbox{\rm Tr}}}
\def\ch{{\hbox{\rm ch}}} \def\imp{{\hbox{\sevenrm Im}}}
\def\trp{{\hbox{\sevenrm Tr}}} \def\vol{{\hbox{\rm Vol}}}
\def\rl{\Lambda_{\hbox{\sevenrm R}}} \def\wl{\Lambda_{\hbox{\sevenrm W}}}
\def\fc{{\cal F}_{k+\cox}} \def\vev{vacuum expectation value}
\def\nodiv{\mid{\hbox{\hskip-7.8pt/}}}
\def\ie{{\em i.e.}}
\def\ie{\hbox{\it i.e.}}

\def\CC{{\mathchoice
{\rm C\mkern-8mu\vrule height1.45ex depth-.05ex
width.05em\mkern9mu\kern-.05em}
{\rm C\mkern-8mu\vrule height1.45ex depth-.05ex
width.05em\mkern9mu\kern-.05em}
{\rm C\mkern-8mu\vrule height1ex depth-.07ex
width.035em\mkern9mu\kern-.035em}
{\rm C\mkern-8mu\vrule height.65ex depth-.1ex
width.025em\mkern8mu\kern-.025em}}}

\def\RR{{\rm I\kern-1.6pt {\rm R}}}
\def\NN{{\rm I\!N}}
\def\ZZ{{\rm Z}\kern-3.8pt {\rm Z} \kern2pt}
\def\IB{\relax{\rm I\kern-.18em B}}
\def\ID{\relax{\rm I\kern-.18em D}}
\def\II{\relax{\rm I\kern-.18em I}}
\def\IP{\relax{\rm I\kern-.18em P}}
\newcommand{\CS}{{\scriptstyle {\rm CS}}}
\newcommand{\CSs}{{\scriptscriptstyle {\rm CS}}}
\newcommand{\rc}{\nonumber\\}
\newcommand{\bear}{\begin{eqnarray}}
\newcommand{\eear}{\end{eqnarray}}

\newcommand{\LL}{{\cal L}}

\def\mani{{\cal M}}
\def\calo{{\cal O}}
\def\calb{{\cal B}}
\def\calw{{\cal W}}
\def\calz{{\cal Z}}
\def\cald{{\cal D}}
\def\calc{{\cal C}}
\def\to{\rightarrow}
\def\ele{{\hbox{\sevenrm L}}}
\def\ere{{\hbox{\sevenrm R}}}
\def\zb{{\bar z}}
\def\wb{{\bar w}}
\def\nodiv{\mid{\hbox{\hskip-7.8pt/}}}
\def\menos{\hbox{\hskip-2.9pt}}
\def\dr{\dot R_}
\def\drr{\dot r_}
\def\ds{\dot s_}
\def\da{\dot A_}
\def\dga{\dot \gamma_}
\def\ga{\gamma_}
\def\dal{\dot\alpha_}
\def\al{\alpha_}
\def\cl{{closed}}
\def\cls{{closing}}
\def\vev{vacuum expectation value}
\def\tr{{\rm Tr}}
\def\to{\rightarrow}
\def\too{\longrightarrow}


\def\a{\alpha}
\def\b{\beta}
\def\c{\gamma}
\def\d{\delta}
\def\e{\epsilon}           
\def\F{\Phi}
\def\f{\phi}               
\def\vf{\varphi}  \def\tvf{\tilde{\varphi}}
\def\vp{\varphi}
\def\g{\gamma}
\def\h{\eta}
\def\j{\psi}
\def\k{\kappa}                    
\def\l{\lambda}
\def\m{\mu}
\def\n{\nu}
\def\o{\omega}  \def\w{\omega}
\def\q{\theta}  \def\th{\theta}                  
\def\r{\rho}                                     
\def\s{\sigma}                                   
\def\t{\tau}
\def\u{\upsilon}
\def\x{\xi}
\def\X{\Xi}
\def\z{\zeta}
\def\pt{\tilde{\varphi}}
\def\tt{\tilde{\theta}}
\def\lab{\label}
\def\6{\partial}
\def\wg{\wedge}
\def\atanh{{\rm arctanh}}
\def\bpsi{\bar{\psi}}
\def\bt{\bar{\theta}}
\def\bvf{\bar{\varphi}}

%

\newfont{\namefont}{cmr10}
\newfont{\addfont}{cmti7 scaled 1440}
\newfont{\boldmathfont}{cmbx10}
\newfont{\headfontb}{cmbx10 scaled 1728}
\newcommand{\re}{\,\mathbb{R}\mbox{e}\,}
\newcommand{\hyph}[1]{$#1$\nobreakdash-\hspace{0pt}}
\providecommand{\abs}[1]{\lvert#1\rvert}
\newcommand{\Nugual}[1]{$\mathcal{N}= #1 $}
\newcommand{\sub}[2]{#1_\text{#2}}
\newcommand{\partfrac}[2]{\frac{\partial #1}{\partial #2}}
\newcommand{\bsp}[1]{\begin{equation} \begin{split} #1 \end{split} \end{equation}}
\newcommand{\calF}{\mathcal{F}}
\newcommand{\calO}{\mathcal{O}}
\newcommand{\calM}{\mathcal{M}}
\newcommand{\calV}{\mathcal{V}}
\newcommand{\bbZ}{\mathbb{Z}}
\newcommand{\bbC}{\mathbb{C}}
\newcommand{\cK}{{\cal K}}
\newcommand{\dd}{\textrm{d}}
\newcommand{\DD}{\textrm{D}}

\newcommand{\Thq}{\Theta\left(\r-\r_q\right)}
\newcommand{\Dq}{\d\left(\r-\r_q\right)}
\newcommand{\kten}{\kappa^2_{\left(10\right)}}
\newcommand{\pbi}[1]{\imath^*\left(#1\right)}
\newcommand{\ho}{\hat{\omega}}
\newcommand{\tth}{\tilde{\th}}
\newcommand{\tf}{\tilde{\f}}
\newcommand{\tj}{\tilde{\j}}
\newcommand{\tw}{\tilde{\omega}}
\newcommand{\tz}{\tilde{z}}
\newcommand{\prj}[2]{(\partial_r{#1})(\partial_{\j}{#2})-(\partial_r{#2})(\partial_{\j}{#1})}
\def\atanh{{\rm arctanh}}
\def\sech{{\rm sech}}
\def\csch{{\rm csch}}
\allowdisplaybreaks[1]

\def\red{\textcolor[rgb]{0.98,0.00,0.00}}

\numberwithin{equation}{section}

\newcommand{\Tr}{\mbox{Tr}}    


%
\renewcommand{\theequation}{{\rm\thesection.\arabic{equation}}}
\begin{titlepage}

\vfill
\begin{flushright}
DMUS-MP-15-08 \\
YITP-SB-15-19 \\
FPAUO-15/07 \\
CAS-KITPC/ITP-451
\end{flushright}

\vfill

\begin{center}
   \baselineskip=16pt
   {\Large \bf  New $AdS_3 \times S^2$ T-duals  with $\mathcal{N} = (0,4)$ supersymmetry}
   \vskip 2cm
     
     Yolanda Lozano$^a$,  Niall T. Macpherson$^b$, Jes\'us Montero$^a$, Eoin \'O Colg\'ain$^{c, d, e}$
       \vskip .6cm
             \begin{small}
                 
                 \textit{$^a$Departamento de F\'isica, 
		 Universidad de Oviedo, 
33007 Oviedo, SPAIN}
                 \vspace{3mm} 
                 
                 \textit{$^b$ Dipartimento di Fisica, Universit\`a di Milano--Bicocca, I-20126 Milano \&
    INFN, sezione di Milano--Bicocca, ITALY}
                 \vspace{3mm} 
                 
                 \textit{$^c$ C.N.Yang Institute for Theoretical Physics, SUNY Stony Brook, NY 11794-3840, USA}
                 \vspace{3mm}
                 
                  \textit{$^d$ Department of Mathematics, University of Surrey, Guildford GU2 7XH, UK}
                 \vspace{3mm}
                 
                  \textit{$^e$ Kavli Institute for Theoretical Physics China, Institute for Theoretical Physics, Chinese Academy of Sciences, Beijing 100190, CHINA}
                 \vspace{3mm}

             \end{small}

     \end{center}

\vfill \begin{center} \textbf{Abstract}\end{center} \begin{quote}

 It is well known that Hopf-fibre T-duality and uplift takes the D1-D5 near-horizon into a class of $AdS_3 \times S^2$ geometries in 11D where the internal space is a Calabi-Yau three-fold. Moreover, supersymmetry dictates that Calabi-Yau is the only permissible $SU(3)$-structure manifold. Generalising this duality chain to non-Abelian isometries, a strong parallel exists, resulting in the first explicit example of a class of $AdS_3 \times S^2$ geometries with $SU(2)$-structure. Furthermore, the non-Abelian T-dual of $AdS_3 \times S^3 \times S^3 \times S^1$ results in a new supersymmetric $AdS_3 \times S^2$ geometry, which falls outside of all known classifications. We explore the basic properties of the holographic duals associated to the new backgrounds. We compute the central charges and show that they are compatible with a large $\mathcal{N}=4$ superconformal algebra in the infra-red. 
  \end{quote} \vfill

\end{titlepage}

\tableofcontents

\setcounter{footnote}{0}
\renewcommand{\theequation}{{\rm\thesection.\arabic{equation}}}

\section{Introduction}
It is not surprising that supersymmetric $AdS_3 \times S^2$ solutions to 11D supergravity \cite{Gauntlett:2006ux, Kim:2007hv} bear a striking resemblance to their $AdS_5 \times S^2$ 
counterparts \cite{Lin:2004nb}; obvious cosmetic differences, such as supersymmetry and G-structures \footnote{Killing spinors, or supersymmetry variations, transform as a doublet under the $SU(2)$ R-symmetry and are tensored with the Killing spinors of AdS$_{d+1}$, which have $2^{\frac{d}{2}}$ complex components.}, are ultimately tied to dimensionality. In common, we note that both spacetimes possess manifest $SU(2)$ isometries, dual to the R-symmetries of the respective 2D $\mathcal{N} = (0,4)$  \cite{Maldacena:1997de, Gaddam:2014mna} and 4D $\mathcal{N} =2$ \cite{Gaiotto:2009we, Gaiotto:2009gz} SCFTs, and that supersymmetric geometries are in one-to-one correspondence with second-order PDEs. For the $\frac{1}{2}$-BPS bubbling geometries of Lin, Lunin and Maldacena (LLM), one famously encounters the 3D continuous Toda equation \cite{Lin:2004nb}, while a similar local analysis in \cite{Kim:2007hv} has revealed a 5D analogue for $\frac{1}{4}$-BPS geometries: 
\be
\label{toda_like}
y \, \partial_{y} \left( y^{-1} \partial_{y} J \right) = \dd_4  \left(  J \cdot \dd_4 \sech^2 \zeta   \right),  
\ee
where the internal space exhibits $SU(2)$-structure \footnote{$SU(2)$-structure in 6D is equivalent to two canonical $SU(3)$-structures.}. Above $\zeta$ is a scalar depending on the 5D coordinates ($y, x_i$), $J$ is the K\"ahler-form of the 4D base and $\dd_4$ denotes the pull-back of the derivative to the base. The 4D base corresponds to an \textit{almost Calabi-Yau} two-fold \cite{Joyce:2001nm}. 

Finding explicit supersymmetric geometries is thus equivalent, at least locally, to solving these PDEs. Despite the difficulties,  we have witnessed a growing number of $AdS_5 \times S^2$ geometries, and associated Toda solutions;  starting with early constructions from gauged supergravity \cite{Maldacena:2000mw}, through examples found directly in 11D \cite{Fayyazuddin:1999zu} \footnote{The 11D solution can be dimensionally reduced and T-dualised, where it becomes a quotient of $AdS_5\times S^5$. This provides no contradiction with a no-go result for $\frac{1}{2}$-BPS AdS$_5$ in IIB Ref. \cite{Colgain:2011hb}.}, recently a large number of solutions have been constructed by exploiting an added isometry and a connection to electrostatics \cite{Gaiotto:2009gz, ReidEdwards:2010qs, Donos:2010va, Aharony:2012tz}.  More recently, exotic solutions without an electrostatic, or with only an emergent electrostatic description have been found \cite{Petropoulos:2013vya, Petropoulos:2014rva}.  Relevant to this current work, it is noteworthy that the $SU(2)$ non-Abelian T-dual of $AdS_5 \times S^5$ also corresponds to a solution in this class \cite{Sfetsos:2010uq}. 

In contrast, little is known about solutions to (\ref{toda_like}). Given the current literature,  if we eliminate geometries exhibiting more supersymmetry, which one can disguise as $AdS_3 ~\times S^2$ (see section 4 of \cite{Kim:2007hv}), there is no known  $\frac{1}{4}$-BPS geometry that solves (\ref{toda_like}). In this paper, after uplift to 11D, we identify the non-Abelian T-dual of $AdS_3 \times S^3 \times CY_2$ \cite{Sfetsos:2010uq} as the first example in this class. Admittedly, this example solves (\ref{toda_like}) in the most trivial way, since $\partial_{y} J = \dd_4 \zeta = 0$. That being said, it should be borne in mind that the linear supersymmetry conditions are satisfied non-trivially. It is worth appreciating an obvious parallel to Abelian T-duality, where the uplifted geometry is an example of an $SU(3)$-structure manifold, namely Calabi-Yau. 

Before proceeding, we touch upon the generality of (\ref{toda_like}). It is not clear if all supersymmetric $\frac{1}{4}$-BPS $AdS_3~ \times S^2$ solutions in 11D with $SU(2)$-structure satisfy (\ref{toda_like}). Indeed, the analysis of LLM made the simplifying assumption that there are no $AdS_5 \times S^2$ geometries with purely magnetic flux. Similarly, \cite{Kim:2007hv} precluded both purely electric and magnetic fluxes \cite{Kim:2007hv}, a choice that is supported by $AdS$-limits of wrapped M5-brane geometries \cite{Gauntlett:2006ux, MacConamhna:2006nb}. For LLM, it can be explicitly shown that extra fluxes are inconsistent with supersymmetry \cite{OColgain:2010ev} \footnote{Generalising the Killing spinor ansatz \cite{OColgain:2012wv} allows one to also describe maximally supersymmetric 11D solutions or $\frac{1}{2}$-BPS pp-waves, such as \cite{Ortiz:2014aja}.}  and an attempt at a more general analysis for $AdS_3 \times S^2$ geometries appeared in \cite{Colgain:2010wb}, which derives the supersymmetry conditions in all generality, but unfortunately fails to constrain the fluxes greatly. Using these conditions, one can show that the existence of a single chiral spinor internally, corresponding to $SU(3)$-structure, implies Calabi-Yau \footnote{A small caveat here is that one of the 6D spinors $\epsilon_{\pm}$ was assumed to be chiral, however the supersymmetry constraints on scalar bilinears are strong enough to ensure $\epsilon_- = -i \epsilon_+$. The Calabi-Yau conditions $\dd J = \dd \Omega = 0$ then follow. We thank D. Tsimpis for raising this loop-hole.}. For $SU(2)$-structure manifolds, we note that the non-Abelian T-dual of $AdS_3 \times S^3 \times CY_2$ fits neatly into the classification of \cite{Kim:2007hv}. In contrast, the non-Abelian T-dual of $AdS_3 \times S^3 \times S^3 \times S^1$ preserves the same supersymmetry, $\mathcal{N} = (0,4)$ in 2D, yet falls outside this class, thus motivating future work to extract the more general class \cite{work_in_progress}. 

Non-Abelian T-duality has revealed itself as a powerful tool to construct explicit $AdS$ solutions that seemed unreachable by other means. In this work we present some further examples.
Interesting solutions generated this way \footnote{See also
\cite{Itsios:2012zv,Sfetsos:2014tza,Macpherson:2014eza,Bea:2015fja} for
further $AdS$ solutions and
\cite{Barranco:2013fza,Macpherson:2013zba,Gaillard:2013vsa,Caceres:2014uoa,Kooner:2014cqa,Zacarias:2014wta,Pradhan:2014zqa,Gevorgyan:2013xka}
for a more varied sample of the NAT duality literature.}
are the only explicit $AdS_6$ solution to Type IIB supergravity constructed in \cite{Lozano:2012au}  \footnote{Supersymmetry imposes severe constraints to the existence of $AdS_6$ solutions in ten and eleven dimensions \cite{Passias:2012vp, Apruzzi:2014qva}. Prior to \cite{Lozano:2012au}  the only known explicit solution to Type II supergravities was the Brandhuber and Oz background \cite{Brandhuber:1999np}, which was shown to be the only possible such solution in (massive) IIA in \cite{Passias:2012vp}. Later \cite{Apruzzi:2014qva} proved the non-existence of $AdS_6$ solutions in M-theory and derived the PDEs that such solutions must satisfy in Type IIB (see also \cite{Kim:2015hya}), to which the example in  \cite{Lozano:2012au}, constructed from the Brandhuber and Oz solution via non-Abelian T-duality, provides the only known explicit solution (besides the Abelian T-dual).} and the recent 
 $\mathcal{N}=2$ $AdS_4$ solution to M-theory with purely magnetic flux constructed in \cite{LMM}, which provides the only such explicit solution besides the Pernici-Sezgin background derived in the eighties \cite{Pernici:1984nw}.
Both these solutions may play an important role as gravity duals of, respectively, 5d fixed point theories arising from Type IIB brane configurations, probably from 7-branes as in \cite{DeWolfe:1999hj} (see \cite{Lozano:2013oma}), and of  3d SCFTs arising from M5-branes wrapped on 3d manifolds in the context of the 3d-3d correspondence \cite{Dimofte:2011ju}. In turn, the new $AdS_3$ backgrounds that we construct in this paper may provide the holographic duals of new  2D large $\mathcal{N}=(0,4)$ field theories arising from D-brane intersections. Other $AdS_3$ backgrounds dual to $\mathcal{N}=(0,2)$ 2D field theories haven been constructed recently in  \cite{Bea:2015fja} (see also
\cite{Araujo:2015npa})
 by compactifying on a 2D manifold the Klebanov-Witten background, combined with Abelian and non-Abelian T-dualities. 

An essential difference with respect to its Abelian counterpart, is that non-Abelian T-duality has not been proved to be a symmetry of String Theory. In the context of the AdS/CFT correspondence one could thus expect new $AdS$ backgrounds from known ones with very different dual CFTs. 
Furthermore, these CFTs are only guaranteed to exist in the strong coupling regime, 
since there is no reason to expect that the transformation should survive $\alpha^\prime$ or $1/N$ corrections. 

Even if the understanding of the CFT interpretation of the transformation is today very preliminary, some results point indeed in these directions. The non-Abelian T-dual of the $AdS_5\times S^5$ background constructed in \cite{Sfetsos:2010uq} has been shown for instance to belong to the family of $\mathcal{N}=2$ Gaiotto-Maldacena geometries \cite{Gaiotto:2009gz}, proposed as duals of the, intrinsically strongly coupled, $T_N$ Gaiotto theories \cite{Gaiotto:2009we}. Similarly, the non-Abelian T-dual of the $AdS_5\times T^{1,1}$ background \cite{Klebanov:1998hh} gives rise to an $AdS_5$ background \cite{Itsios:2013wd}  that belongs to the general class of $\mathcal{N}=1$ solutions in \cite{Bah:2011vv,Bah:2012dg}, whose dual CFTs generalize the so-called Sicilian quivers of \cite{Benini:2009mz}, and  are the $\mathcal{N}=1$ analogues of the $\mathcal{N}=2$ solutions in  \cite{Gaiotto:2009we}.

Some works have tried to explore in more depth the CFT realization of $AdS$ backgrounds generated through non-Abelian T-duality  in different dimensions \cite{Lozano:2013oma}\cite{Itsios:2012zv}-\cite{Bea:2015fja}. Its interplay with supersymmetry  \cite{Kelekci:2014ima} and
phenomenological properties of the dual CFTs, such as the type of branes generating the geometry, the behavior of universal quantities such as the free energy, or the entanglement entropy, the realization of baryon vertices, instantons, giant gravitons, are by now quite systematized (see  \cite{Lozano:2014ata}). Very recently, we have witnessed as well an exciting and novel application in the exchange of particles with vortices \cite{Murugan:2015boa}. 
In this paper we will analyze some of these properties in the 2D holographic duals to the new $AdS_3$ backgrounds that we generate. We will see that they fit in the general picture observed in other dimensions.

Perhaps the most puzzling obstacle towards a precise CFT interpretation of non-Abelian T-duality  is the fact that even if the group used to construct the non-Abelian T-dual background is compact, the original coordinates transforming under this group are replaced in the dual by coordinates living in its Lie algebra. Non-compact internal directions are thus generated, which are hard to interpret in the CFT.
We will also encounter this problem for the backgrounds generated in this paper.

The paper is organized as follows. In section 2 we present the first explicit example of an $AdS_3 \times S^2$ geometry belonging to the general class of solutions \cite{Kim:2007hv}. This is constructed by uplifting the non-Abelian T-dual of $AdS_3\times S^3\times CY_2$ derived in \cite{Sfetsos:2010uq} to 11D. In section 3 we recall the basic properties of the $AdS_3\times S^3\times S^3\times S^1$ background that will be the basis of the new solutions that we present in sections 4, 5 and 6. In section 4 we construct the non-Abelian T-dual of this background with respect to a freely acting $SU(2)$ on one of the $S^3$. By exploring the solution we derive some properties of the associated dual CFT such as the central charge and the 
type of color and flavor branes from which it may arise. We suggest a possible explicit realization in terms of intersecting branes. In section 5 we construct one further solution through Abelian T-duality plus uplift to 11D from the previous one and show that it provides an explicit example of an $AdS_3\times S^2$ geometry in 11D belonging to a new class that is beyond the ansatz in \cite{Kim:2007hv}. In section 6 we present a new $AdS_3 \times S^2\times S^2$ solution to Type IIB obtained by further dualizing the solution in section 3 with respect to a freely acting $SU(2)$ on the remaining $S^3$.
By analyzing the same brane configurations we argue that the field theory dual shares some common properties with the CFT dual to the original $AdS_3\times S^3\times S^3\times S^1$ background but in a less symmetric fashion. 
In section 7 we analyze in detail the supersymmetries preserved by the different solutions that we construct. We show that the solutions constructed through non-Abelian T-duality from the $AdS_3\times S^3\times S^3\times S^1$ background exhibit large $\mathcal{N}=(0,4)$ supersymmetry. This is supported by the analysis of the central charges performed in sections 4 and 6. Section 8 contains our conclusions. Finally, in the Appendix we study in detail the effect of Hopf-fibre T-duality in the $AdS_3\times S^3\times S^3\times S^1$ background to further support our claims in the text concerning the isometry supergroup of our solutions.

\section{$AdS_3 \times S^2$ geometries in 11D with $SU(2)$-structure}
\label{sec:SU2structure}

In this section we demonstrate that the non-Abelian T-dual of the D1-D5 near-horizon, a solution that was originally written down in \cite{Sfetsos:2010uq}, uplifts to 11D, where it provides the first explicit example of a $\frac{1}{4}$-BPS $AdS_3 \times S^2$ geometry with an internal $SU(2)$-structure manifold. We recall that this class has appeared in a series of classifications \cite{Gauntlett:2006ux, Kim:2007hv, MacConamhna:2006nb, Colgain:2010wb}, yet until now, not a single explicit example in this class was known. It is indeed pleasing to recognise that the chain of dualities that generates this new example is no more than a simple non-Abelian generalisation of a well-known mapping from the $AdS_3 \times S^3 \times CY_2$ geometry of Type IIB supergravity into the 11D supergravity class $AdS_3 \times S^2 \times CY_3$ \footnote{See appendix B of ref. \cite{Karndumri:2013dca} for a concrete realisation of the (Abelian) duality chain.}. It is worth noting that until relatively recently \cite{Sfetsos:2010uq} (also \cite{Itsios:2013wd}), the workings of this new mapping, which is made possible through non-Abelian T-duality, were also unknown. 

We begin by reviewing the classification of ref. \cite{Colgain:2010wb}, which has an advantage over other approaches \cite{Gauntlett:2006ux}, since it uses local techniques and is thus guaranteed to capture all supersymmetric solutions. Moreover, this work also extends the ansatz of ref. \cite{Kim:2007hv} and dispenses with the need for an analytic continuation from $S^3 \times S^2$ to $AdS_3 \times S^2$. Based on symmetries, the general form for a supersymmetric spacetime of this type may be expressed as  
\bea
\label{gen_ansatz}
\dd s^2_{11} &=& e^{2 \lambda} \left[ \frac{1}{m^2} \dd s^2 (AdS_3) + e^{2 A} \dd s^2 (S^2) + \dd s_6^2 \right],  \cr
G_4 &=& \textrm{Vol} (AdS_3) \wedge \mathcal{A} +  \textrm{Vol} (S^2) \wedge \mathcal{H} + \mathcal{G}, 
\eea
where $\lambda, A$ denote warp-factors depending on the coordinates of the 6D internal space and $\mathcal{A}, \mathcal{H}$ and $\mathcal{G}$ correspond to one, two and four-forms, respectively, with legs on the internal space. The constant $m$ denotes the inverse radius of $AdS_3$. The supersymmetry conditions, which are given in terms of differential conditions on spinor bilinears, further built from two a priori independent 6D spinors $\epsilon_{\pm}$, can be found in \cite{Colgain:2010wb}. 

Setting $\mathcal{A} = \mathcal{G} = 0$, one finds that only a particular linear combination, $\tilde{\epsilon} = \epsilon_{+} \pm i \gamma_7 \epsilon_-$ appears in the effective 6D Killing spinor equations, allowing one to recover the work of \cite{Kim:2007hv}. In this simplifying case one can show that the internal space must be of the form \cite{Kim:2007hv, Colgain:2010wb}
\be
\dd s^2_6 = g_{ij} \dd x^i \dd x^j + e^{-6 \lambda} \sec^2 \zeta \dd y^2 + \cos^2 \zeta  (\dd \psi + P)^2 
\ee
with $P$ a one-form connection on the 4D base with metric $g_{ij}$. The $SU(2)$-structure is then specified by 2 one-forms, $K^1 \equiv \cos \zeta ( \dd \psi + P)$, $K^2 \equiv e^{-3 \lambda} \sec \zeta \dd y$, the K\"{a}hler-form, $J$, and the complex two-form, $\Omega$, on the base. 

The remaining two-form appearing in the field strength, $G_4$, is fully determined by supersymmetry, 
\bea
\mathcal{H} &=& - y J - \frac{1}{2 m} \partial_{y} ( y \sin^2 \zeta) \dd y \wedge (\dd \psi + P) - \frac{y}{m} \cos \zeta \sin \zeta \dd_4 \zeta \wedge ( \dd \psi + P)  \cr &+& \frac{y \cos^2 \zeta}{2m} \dd P. 
\eea
The above class of geometries is subject to the supersymmetry conditions: 
\bea
\label{SUSYconds}
2 m y &=& e^{3 \lambda} \sin \zeta, \cr \quad e^{A} &=& \frac{\sin \zeta}{2m},  \cr
\dd ( e^{3 \lambda} \cos \zeta \Omega) &=& 0,  \cr
2 m \dd ( e^{3 \lambda + 2A} J) &=& \dd_4 P \wedge \dd y.
\eea
Details of how (\ref{toda_like}) is implied by these conditions can be found in \cite{Kim:2007hv}. 

In order to identify a solution in this class, we start by recalling the non-Abelian T-dual of $AdS_3 \times S^3 \times T^4$ \cite{Sfetsos:2010uq}, which provides a solution to massive IIA supergravity, 
\bea
\dd s^2_{IIA} &=& \dd s^2 (AdS_3) + \dd \rho^2 + \frac{\rho^2}{1+ \rho^2} \dd s^2 (S^2) + \dd s^2 (T^4), \cr
B_2 &=& \frac{\rho^3}{1+ \rho^2} \textrm{vol} (S^2), \quad \Phi = - \frac{1}{2} \ln (1+ \rho^2), \cr
m &=& 1, \quad F_2 = \frac{\rho^3}{1+ \rho^2} \textrm{vol}(S^2), \cr
F_4 &=&\textrm{vol} (AdS_3) \wedge \rho \dd \rho + \textrm{vol}(T^4),  
\eea
where following \cite{Sfetsos:2010uq}, we have suppressed factors associated to radii for simplicity. As a consequence, the $AdS_3$ metric is normalised so that $R_{\mu \nu} = -\frac{1}{2} g_{\mu \nu}$, whereas $S^2$ is canonically normalised to unit radius. 

We next perform two T-dualities along the $T^4$, the coordinates of which we label, $x_1, \dots x_4$. Performing T-dualities with respect to $x_1$ and $x_2$, we can replace the Romans' mass, $m=1$, with higher-dimensional forms, while leaving the NS sector unaltered. In addition to the NS two-form, the geometry is then supported by the following potentials from the RR sector,  
\bea
C_1 &=& \frac{1}{2} ( x_1 \dd x_2 - x_2 \dd x_1 + x_3 \dd x_4 - x_4 \dd x_3), \cr
C_3 &=& \frac{\rho^3}{1+ \rho^2} \textrm{vol} (S^2) \wedge C_1.  
\eea 
We note that $\dd C_1 = J$, where $J$ is the K\"ahler form on $T^4$ and the Bianchi for $F_4$, namely $\dd F_4 = H_3 \wedge F_2$ is satisfied in a trivial way since $F_4 = \dd C_3 = B_2 \wedge J$. We can now uplift the solution on a circle to 11D by considering the standard Kaluza-Klein ansatz, 
\bea
\label{11D_uplift}
\dd s^2_{11} &=& (1+ \rho^2)^{\frac{1}{3}} \left[ \dd s^2(AdS_3) + \frac{\rho^2}{1+ \rho^2} \dd s^2(S^2) + \dd \rho^2 + \dd s^2(T^4) \right] + (1+ \rho^2)^{-\frac{2}{3}} \DD z^2, \cr
G_4 &=& \textrm{vol} (S^2) \left[ \frac{\rho^3}{1+ \rho^2} J + \frac{\rho^2 ( \rho^2 + 3)}{(1+ \rho^2)^2} \dd \rho  \wedge \DD z\right], 
\eea
where we have defined $\DD z \equiv \dd z + C_1$. 

Adopting $m=2$, so that normalisations for $AdS_3$ agree, and up to an overall sign in $\mathcal{H}$, which can be accommodated through the sign flip $\rho \leftrightarrow - \rho$, we find that the supersymmetry conditions (\ref{SUSYconds}) are satisfied once one identifies
accordingly
\bea
y &=& \rho, \quad e^{\lambda} = (1+ \rho^2)^{\frac{1}{6}}, \quad e^{A} = \frac{\rho}{(1+ \rho^2)^{\frac{1}{2}}}, \quad P = C_1, \cr
J &=& \dd x_1 \wedge \dd x_2 + \dd x_3 \wedge \dd x_4, \cr
\Omega &=& (\dd x_1 + i \dd x_2) \wedge ( \dd x_3 + i \dd x_4). 
\eea
Thus the non-Abelian T-dual plus 11D uplift of the D1-D5 near horizon fits in the classifications \cite{Gauntlett:2006ux, Kim:2007hv,  MacConamhna:2006nb, Colgain:2010wb}.
It is easy to see that one can replace $T^4$ with K3 and the construction still holds. It is also easy to see that the above solution can be derived on the assumption that the base is Calabi-Yau and that $\lambda, \zeta$ only depend on $y$. Indeed, this is a requirement of the 6D $SU(2)$-structure manifold to be a complex manifold \cite{Colgain:2010wb}. In this case, the supersymmetry conditions imply $e^{3 \lambda} \cos \zeta$ is a constant. We can then solve for $\lambda, \zeta$ and $A$ giving us the above solution. 


Another interesting feature of the 11D solution is that in performing the classification exercise using Killing spinor bilinears \cite{Kim:2007hv, Colgain:2010wb}, one finds a $U(1)$ isometry that emerges from the analysis for free. Often this $U(1)$ corresponds to an R-symmetry, for example \cite{Lin:2004nb, Gauntlett:2004zh}, but in this setting, the relevant superconformal symmetry in 2D either corresponds to small superconformal symmetry with R-symmetry $SU(2)$, or large superconformal symmetry with R-symmetry $SU(2) \times SU(2)$. There appears to be no place for a $U(1)$ R-symmetry and it is an interesting feature of solutions fitting into the class of \cite{Kim:2007hv} that the $U(1)$ is the M-theory circle and the Killing spinors are uncharged with respect to this direction \footnote{In 11D one can identify the two projection conditions to verify that supersymmetry is not enhanced. From $CY_2$ directions, we inherit $\Gamma^{6789} \eta = - \eta$, the rotation on the 11D spinor becomes $\eta = \exp [-\frac{1}{2} \tan^{-1} \left( \frac{1}{\rho} \right) \Gamma^{\chi \xi z} ] \tilde{\eta}$. One finds the additional projector, $\Gamma^{\rho z 67} \tilde{\eta} = - \tilde{\eta}$, thus confirming that the 11D solution is indeed $\frac{1}{4}$-BPS.}. 

In the rest of this paper, we study non-Abelian T-duals of another well-known $\frac{1}{2}$-BPS $AdS_3$ solution with $\mathcal{N} = (4,4)$ supersymmetry, namely $AdS_3 \times S^3 \times S^3 \times S^1$, where we will find a new supersymmetric solution that does not fit into the class in \cite{Kim:2007hv}.

\section{The $AdS_3\times S^3\times S^3\times S^1$ background with pure RR flux}

In this section we recall the basic properties of the $AdS_3\times S^3\times S^3\times S^1$  background \cite{Cowdall:1998bu}-\cite{Gukov:2004ym}, which will be the basis of our study in the following sections.

The  $AdS_3\times S^3\times S^3\times S^1$ background is a half-BPS solution of Type II string theory supported by NS5-brane and string flux. In this paper we will be interested in its realization in Type IIB where it is supported by D5 and D1-brane fluxes \cite{de Boer:1999rh}. This description arises after compactifying on a circle the $AdS_3\times S^3\times S^3\times \mathbb{R}$ near horizon geometry of a D1-D5-D5' system where the two stacks of D5-branes are orthogonal and intersect only along the line of the D1-branes \cite{Cowdall:1998bu,Boonstra:1998yu,Gauntlett:1998kc}. How to implement the $S^1$ compactification has remained unclear (see \cite{de Boer:1999rh}), and it has only been argued recently \cite{Tong:2014yna} that the $\mathbb{R}$  instead of the $S^1$ factor arising in the near horizon limit  could just be an artefact of the smearing of the D1-branes on the transverse directions prior to taking the limit. This reference has also provided the explicit $\mathcal{N}=(4,4)$ CFT realization conjectured in \cite{Elitzur:1998mm,de Boer:1999rh,Gukov:2004ym} for the field theory dual. This CFT arises as the infrared fixed point of the $\mathcal{N}=(0,4)$ gauge theory living on the D1-D5-D5' intersecting D-branes.

The $AdS_3\times S^3_+\times S^3_-\times \mathbb{R}$ metric is given by 
\beq
\label{original}
ds^2_{IIB}= L^2 \dd s^2 (AdS_3)+ R_+^2 \dd s^2 (S^3_+)+R_-^2 \dd s^2 (S^3_-)+\dd x^2
\eeq
with
\begin{equation}
\dd s^2 (AdS_3)=r^2 (-\dd t^2+\dd x_1^2)+\frac{\dd r^2}{r^2}
\end{equation}
in Poincar\'e coordinates. Plus, the background is supported by a single non trivial RR flux
\beq
F_3= 2 L^2 \textrm{Vol}(AdS_3)+ 2R_+^2 \textrm{Vol}(S^3_+)+ 2R_-^2 \textrm{Vol}(S^3_-),
\eeq
with Hodge dual
\begin{equation}\label{eq:F7orig}
 F_7=  \Big\{ 2 L^3 \textrm{Vol}(AdS_3)\wedge\Big( - \frac{R_+^3}{R_-}  \textrm{Vol}(S^3_+) +  \frac{R_-^3}{R_+} \textrm{Vol}(S^3_-) \Big)  +  \frac{2 R_+^3 R_-^3}{L} \textrm{Vol}(S^3_+)\wedge \textrm{Vol}(S^3_-)  \Big\}\wedge \dd x   
\end{equation}
We take $g_s=1$ such that the dilaton is zero and Einstein's equations are satisfied only when
\beq
\label{Lerres}
\frac{1}{L^2}=\frac{1}{R_+^2}+\frac{1}{R_-^2}.
\eeq

This background has a large invariance under $SO(4)^+\times SO(4)^-$ spatial rotations. Of these $SU(2)^+_R\times SU(2)^-_R$ correspond to the R-symmetry group of the $\mathcal{N}=(0,4)$ field theory living at the D1-D5-D5' intersection, and $SU(2)^+_L\times SU(2)^-_L$ to a global symmetry. The field theory has gauge group $U(N_1)$, with $N_1$ the number of D1-branes, and a global symmetry $SU(N_5^+)\times SU(N_5^-)$, with $N_5^+$ and $N_5^-$ the number of D5 and D5' branes.  
The two R-symmetries give rise to two current algebras at levels depending on the background charges, and to a large $\mathcal{N}=(4,4)$ superconformal symmetry in the infra-red \cite{Elitzur:1998mm,de Boer:1999rh,Gukov:2004ym,Tong:2014yna}.

The study of the supergravity solution allows to derive properties of the dual field theory that we will be able to mimic after the non-Abelian T-duality transformation. In the next subsections we analyze the quantized charges, some brane configurations such as baryon vertices and 't Hooft monopoles, and the central charge associated to the $AdS_3\times S^3\times S^3\times S^1$ background.

\subsection{Quantized charges}

The $F_7$ and $F_3$ fluxes generate D1 and  D5-brane charges given by:
\begin{equation}
\label{eq:N1orig}
   N_1= \frac{1}{2\pi \kappa_{10}^2 T_1} \int (-F_7) =   \frac{R_+^3 R_-^3 \delta x}{8 L \pi^2} \;,
\end{equation}
where $\delta x$ is the length of the $x$-direction interval, which should be chosen such that $N_1$ is quantized, and
\begin{equation}
N_5^+ = \frac{1}{2\pi \kappa_{10}^2 T_5} \int_{S^3_-} (-F_3) =  R_-^2 \, , \qquad
N_5^- = \frac{1}{2\pi \kappa_{10}^2 T_5} \int_{S^3_+} (-F_3) =  R_+^2 \, ,
\end{equation}
which should also be quantized. Accordingly, one can find D1 and D5 BPS solutions. The D1 are extended along the $\{t,x_1\}$ directions and couple to the potential
\begin{equation}
C_2= L^2 r^2 \dd t\wedge \dd x_1\, .
\end{equation}
\noindent Changing coordinates to 
\begin{equation}
\label{coordinates-inter}
  \begin{dcases}
   r &= r_+ \, r_- \\
   x &= \frac{R_+^2}{\sqrt{R_+^2+R_-^2}} \log r_+  - \frac{R_-^2}{\sqrt{R_+^2+R_-^2}} \log r_-\, ,
  \end{dcases}
\end{equation} 
the metric becomes the near horizon limit of the intersecting D1-D5-D5' configuration \cite{Cowdall:1998bu,Boonstra:1998yu,Gauntlett:1998kc}:
\begin{eqnarray}
&&N_5^+ \,\,D5: \qquad 012345 \nonumber\\
&&N_5^- \,\,D5^\prime: \quad\,\,\,\, 016789 \nonumber \\
&&N_1 \, D1: \qquad \, \,\, 01
\end{eqnarray}
with
$\dd x_2^2+\dots +\dd x_5^2=\dd r_+^2+r_+^2 \dd s^2(S^3_+)$, 
$\dd x_6^2+\dots +\dd x_9^2=\dd r_-^2+r_-^2 \dd s^2(S^3_-)$, with the D1-branes smeared on these directions:
\beq
\dd s^2_{IIB}= L^2 r_+^2 r_-^2 (-\dd t^2+\dd x_1^2)+ R_+^2\, \frac{\dd r_+^2}{r_+^2} + R_-^2\, \frac{\dd r_-^2}{r_-^2}  + R_+^2 \dd s^2(S^3_+)+R_-^2 \dd s^2 (S^3_-) \;.
\eeq
\noindent The BPS D5-branes are then found lying on the $(t, x_1, r_+, S^3_+)$, $(t, x_1, r_-, S^3_-)$ directions. 

The 2D  $\mathcal{N}=(0,4)$ gauge theory living on the worldvolume of the D1-branes and intersecting D5-branes has been identified recently in \cite{Tong:2014yna}. A key role is played by the chiral fermions of the D5-D5' strings that lie at the intersection. Quite remarkably the central charge of the $\mathcal{N}=(4,4)$ CFT to which this theory flows in the infra-red has been shown to coincide with the central charge of the supergravity solution, that we review in subsection 3.4.

\subsection{Instantons}

The previous configuration of D5, D5' branes joined in a single manifold, where the D1-branes lie, admits a Higgs branch where the D1-branes are realized as instantons in the D5-branes \cite{Gukov:2004ym}. One can indeed compute the quadratic fluctuations of the D5-branes to obtain the effective YM coupling:
\begin{equation}
S^{D5}_{\rm fluc}=-\int \frac{1}{g_{D5}^2} F^2_{\mu\nu} \qquad {\rm with} \qquad \frac{1}{g_{D5}^2}=\frac{L^2 r_+^2 r_-^2}{4(2\pi)^3}
\end{equation}
and check that the DBI action of the D1-branes satisfies
\begin{equation}
S_{DBI}^{D1}=-\int\frac{16\pi^2}{g_{D5}^2}
\end{equation}
as expected for an instantonic brane. 

\subsection{Baryon vertices and 't Hooft monopoles}

A D7-brane wrapped on $S^3_+ \times S^3_- \times S^1$ realizes a baryon vertex in the $AdS_3\times S^3_+\times S^3_-\times S^1$ geometry, since it develops a tadpole of $N_1$ units, as it is inferred from its CS action:
\begin{equation}
 S_{CS}^{D7}=  2\pi\, T_7 \int  C_6\wedge F =-2\pi\, T_7 \int_{S^3_+ \times S^3_- \times S^1}\! \!\!\! F_7 \int \dd t A_t= -N_1 \int \dd t A_t \;,
\end{equation}
where $\delta x$ is taken to satisfy that $N_1$ is an integer as in (\ref{eq:N1orig}). 
 
Similarly, there are two t'Hooft monopoles associated to the ranks of the two flavor groups that are realized in the $AdS_3\times S^3_+\times S^3_-\times S^1$ background as D3-branes wrapping the $S^3_\pm$. The corresponding Chern-Simons terms show that these branes have tadpoles of $N_5^\mp$ units that should be cancelled with the addition of these numbers of fundamental strings:
\begin{equation}
 S_{CS}^{D3^\pm}=  -2\pi T_3 \int_{S^3_\pm} F_3 \int \dd t A_t=  N_5^\mp \int \dd t A_t \;.
\end{equation}

\subsection{Central charge}

The central charge associated to the $AdS_3\times S^3_+\times S^3_-\times S^1$ background can be computed using the Brown-Henneaux formula \cite{Brown:1986nw}, giving \cite{de Boer:1999rh,Gukov:2004ym}:
\begin{equation}
\label{centralc}
c=2N_1\,\frac{N_5^+ N_5^-}{N_5^++N_5^-}\, .
\end{equation}
This expression agrees with the central charge for a large $\mathcal{N}=(4,4)$ CFT with affine $SU(2)^\pm$ current algebras at levels $k^\pm$:
$c=2k^+ k^-/(k^++k^-)$ \cite{Sevrin:1988ew}, with $k^\pm=N_1 N_5^\pm$. A strong check of the validity of the $\mathcal{N}=(0,4)$ field theory  in the D1-branes proposed in \cite{Tong:2014yna} is that it correctly reproduces  (\ref{centralc}) at the infrared fixed point (see also
\cite{Gukov:2004ym}).

\section{Non-Abelian T-dual $AdS_3\times S^3\times S^2$ solution in IIA}
\label{sec:massiveIIA}

In this section we dualize the $AdS_3\times S^3_+\times S^3_-\times S^1$ solution with respect to the $SU(2)^-_L$ acting on the $S^3_-$. This dualization was reported in \cite{Kelekci:2014ima} to produce a new $AdS_3$ solution preserving 16 supercharges. As we shall demonstrate in section \ref{sec:SUSY},  \cite{Kelekci:2014ima} overlooked an extra implied condition and the preserved supersymmetry is in fact 8 supercharges. The solution thus preserves large $\mathcal{N}=(0,4)$ supersymmetry in 2D. In this section we present a detailed study of the geometry and infer some properties about the field theory interpretation of this solution.

\subsection{Background}

Applying the general rules in \cite{Itsios:2012dc} (see also  \cite{Kelekci:2014ima}) we find a dual metric
\beq\label{eq:dual_metric}
\dd s^2_{IIA}= L^2 \dd s^2 (AdS_3)+ R_+^2 \dd s^2(S^3_+)+\frac{4}{R_-^2}\bigg( \dd \rho^2+\frac{R_-^6 \rho^2}{64\Delta}\big(\dd \chi^2+\sin^2\chi \dd\xi^2\big)\bigg)+\dd x^2,
\eeq
where
\beq \label{eq:detM}
\Delta= \frac{R_-^6+16R_-^2\rho^2}{64}.
\eeq
The dual dilaton is given by
\beq  \label{eq:dilaton}
e^{-2\Phi}=\Delta,
\eeq
while the NS 2-form is simply
\beq
\label{NSNS2form}
B_2= \frac{R_-^2 \rho^3}{4 \Delta} {\rm Vol}(S^2)
\eeq
where $S^2$ refers to the 2-sphere parametrised by $0\leq\chi\leq \pi,~ 0\leq \xi<2\pi$ in 
(\ref{eq:dual_metric}).

The dual RR-sector is given by
\beq
\label{dual_RR}
\begin{array}{ll}
m&=\frac{R_-^2}{4},\\[4 mm]
\hat{F}_2&=0,\\[4 mm]
\hat{F}_4&=-\frac{R_-^3}{4L R_+}\big(L^4 {\rm Vol}(AdS_3)+R_+^4 {\rm Vol}(S^3_+)\big)\wedge \dd x + 2 \rho\big(L^2 {\rm Vol}(AdS_3)+R_+^2 {\rm Vol}(S^3_+)\big)\wedge \dd \rho,\\[4 mm]
\hat{F}_6&=-2 L^2 \rho^2 {\rm Vol}(AdS_3)\wedge {\rm Vol}(S^2)\wedge \dd \rho- 2 R_+^2 \rho^2 {\rm Vol}(S_+^3)\wedge {\rm Vol}(S^2)\wedge \dd \rho,\\ [4 mm]
\hat{F}_8&=  -\frac{2 L^3 R_+^3\rho}{R_-}{\rm Vol}(AdS_3)\wedge {\rm Vol}(S^3_+)\wedge \dd x\wedge \dd \rho, \\ [4 mm]
\hat{F}_{10}&= \frac{2L^3 R_+^3 \rho^2}{R_-}{\rm Vol}(AdS_3)\wedge {\rm Vol}(S^3_+) \wedge {\rm Vol}(S^2)\wedge \dd x\wedge \dd \rho.\\ [4 mm]
\end{array}
\eeq
Here ${\hat F}=F e^{-B_2}$ and $F_p=\dd C_{p-1}-H_3\wedge C_{p-3}$. Page charges will be computed from  these ${\hat F}$ according to $\dd * {\hat F}=*j^{\rm Page}$.

Applying the results in \cite{Itsios:2012dc} this background is guaranteed to satisfy the (massive) IIA supergravity equations of motion. Given that the $S^3$ on which we have dualized has constant radius the non-Abelian T-dual solution is also automatically non-singular. An open problem though is the range of the new coordinate $\rho$, which as a result of the non-Abelian T-duality transformation lives in $\mathbb{R}^+$. 

The generation of non-compact directions under non-Abelian T-duality is indeed a generic feature that does not occur under its Abelian counterpart. In the last case the extension of the transformation beyond tree level in string perturbation theory determines uniquely the global properties of the, in principle non-compact, coordinate that replaces the dualized U(1) direction. How to extend non-Abelian T-duality beyond tree level is however a long standing open problem (see \cite{Alvarez:1993qi} for more details), and as a result we are lacking a general mechanism that allows to compactify the new coordinates.  For freely acting SU(2) examples we need to account in particular for the presence of the non-compact $\rho$-direction in the dual internal geometry, which poses a problem to its CFT interpretation, where we can expect operators with continuous conformal dimensions. Note that in the  $AdS_3\times S^2\times S^1$ duals under consideration in this paper one cannot hope that the same mechanism that should be at work for compactifying the $\mathbb{R}$ factor arising in the original $AdS_3\times S^3\times S^3\times \mathbb{R}$ geometry should be applicable. As argued in \cite{Tong:2014yna}, the $\mathbb{R}$ instead of the $S^1$ factor arising in the near horizon limit could be due to the smearing of the D1-branes on the transverse directions, and could thus be avoided with a supergravity solution describing localized branes. This is not directly applicable to our situation because $\rho$ is not an isometric direction.

Previous approaches in the recent non-Abelian T-duality literature have tried to infer global properties through imposing consistency to the dual CFT \cite{Lozano:2013oma,Lozano:2014ata}. We will also follow this approach in this paper. We should start noticing that the new $AdS_3$ metric described by (\ref{eq:dual_metric}) is perfectly regular for all $\rho\in [0,\infty)$, with the 3d space replacing the $S^3_-$ in the original background becoming 
$\mathbb{R}^3$ for small $\rho$ and $\mathbb{R}\times S^2$ for large $\rho$. 
As shown in \cite{Lozano:2013oma,Lozano:2014ata} the definition of large gauge transformations in the dual geometry can give however non-trivial information about its global properties.

\subsection{Large gauge transformations}

The relevance of large gauge transformations is linked to the existence of non-trivial 2-cycles in the geometry, where 
\begin{equation}
\label{largegauge}
	 \frac{1}{4\pi^2} \left| \int_{{\rm 2-cycle}} B_2 \right| \in [0,1)\, .
\end{equation}

In our non-singular metric we can only guarantee the existence of a non-trivial $S^2$ for large $\rho$.  
For finite $\rho$ and given the absence of any global information, we will resort to the most general situation in which the cycle remains non-trivial and we need to care about large gauge transformations. We will see that consistency of the CFT in this most general situation will lead to the condition of vanishing large gauge transformations, which is compatible with the original situation in which the two-cycle may in fact be trivial at finite $\rho$. 

Assuming the existence of a non-trivial two-cycle at finite $\rho$, the $\rho$ dependence of $B_2$ in (\ref{NSNS2form}) implies that large gauge transformations must be defined such that (\ref{largegauge}) is satisfied as we move in this direction. This implies that for $\rho\in [\rho_n,\rho_{n+1}]$ with $\rho_n$ determined by $16\rho_n^3/(R_-^4+16\rho_n^2)=n\pi$,
$B_2$ must be given by
\begin{equation}
\label{B2n}
B_2= \Bigl(\frac{R_-^2 \rho^3}{4 \Delta} -n\pi\Bigr) {\rm Vol}(S^2)\, .
\end{equation}

The fluxes from which the Page charges are computed then change in the different intervals to
\begin{equation}
 \begin{array}{ll}
 \hat{F}_2&\rightarrow  \hspace{0.1cm} \hat{F}_2 + n \pi \, \hat{F}_0 \,  {\rm Vol}(S^2)\\[4 mm]
 \hat{F}_6&\rightarrow \hspace{0.1cm}  \hat F_6 + n \pi \,  \hat{F}_4 \wedge {\rm Vol}(S^2)\, ,  \\ [4 mm]
  \end{array}
\end{equation}
which will affect the values of the Page charges that we compute next.

\subsection{Quantized charges}

The transformation of the RR fluxes under non-Abelian T-duality implies that the D1 color branes of the original background transform into D2-branes extended on $\{t, x_1, \rho\}$ and D4-branes on $\{t, x_1, \rho, S^2\}$. Analogously, the D5 flavor branes wrapped on the $S^3_-$ are mapped into D2 and D4 branes
wrapped on $\{t, x_1, r_-\}$ and $\{t, x_1, r_-, S^2\}$ respectively,
and the D5 transverse to the $S^3_-$ are transformed into D6 and D8 branes wrapped on $\{t, x_1,r_+, S^3_+, \rho\}$ and $\{t, x_1, r_+, S^3_+, \rho, S^2\}$, respectively.
We show in this section that there are quantized charges in the non-Abelian T-dual background that can be associated to these branes.

\subsubsection{Color branes}

It is possible to define $N_2$ and $N_4$ quantized charges in the dual background that should be associated to D2 and D4 color branes:
\begin{align}\label{eq:N1_2_4}
N_{4} &= \frac{1}{2\pi \kappa_{10}^2 T_4} \int_{S_+^3\times S^1} \hat F_4 =  \frac{R_+^3 R_-^3}{16 \pi L}\, \delta x \;, \\ \label{second}
N_{2} &= \frac{1}{2\pi \kappa_{10}^2 T_2} \int_{S_+^3\times S^2\times S^1} \hat F_6 =  nN_4 \;, 
\end{align}
where $n$ is the parameter labeling large gauge transformations. 
(\ref{second}) is the value of the D2 charge
in the $\rho\in [\rho_n, \rho_{n+1}]$ interval. Note that, as it seems to be quite generic under non-Abelian T-duality, the condition imposed on the geometry by  (\ref{eq:N1_2_4}) is different, and in fact incompatible, from the one that the original background satisfied, given by (\ref{eq:N1orig}).  A re-quantization must thus be done in the new background.

Let us now analyze the condition (\ref{second}). We first see that for zero $n$ the charge associated to the D2-branes vanishes.
Second, as we change interval, $N_2$ undergoes a transformation, $N_2\rightarrow N_2-N_4$, that is very reminiscent of Seiberg duality \cite{Seiberg:1994pq}. This was proposed in \cite{Lozano:2014ata} as a way to relate the CFTs dual to the solution as we move in $\rho$. As stressed in \cite{Bea:2015fja}\footnote{We would also like to acknowledge fruitful conversations with D. Rodr\'{\i}guez-G\'omez on this issue.} this cannot be however the full story since there is a change in the number of degrees of freedom as we move in $\rho$. This is explicit in the holographic free energies. The precise realization in the CFT of the running of $\rho$ remains at the very heart of our full understanding of the interplay between non-Abelian T-duality and AdS/CFT. We hope we will be able to report progress in this direction in future publications.

For the particular background considered in this paper it is only possible to find BPS color and flavor branes when $n=0$. In particular, color branes are D4-branes wrapped on the $\{t, x_1, \rho, S^2\}$ directions. Thus, we will take the view that $\rho$ is restricted to the fundamental region $[0,\rho_1]$, with $\rho_1$ satisfying $16\rho_1^3/(R_-^4+16\rho_1^2)=\pi$. Choosing to end the geometry at a regular point presents however other problems, now for the geometry, where extra localized sources should be included. It was proposed in \cite{Bea:2015fja} that at these transition points new gauge groups would be added to the CFT through an ``unhiggsing" mechanism not associated to an energy scale. Given that this mechanism relies in the existence of large gauge transformations it does not seem applicable to our background. A full understanding of the ``unhiggsing'' mechanism and its precise realization in the absence of an energy scale remains as an interesting open problem.

\subsubsection{Flavor branes}

Let us now examine  flavor branes in the dual background. 
We find the following quantized charges in the dual background that should be associated to flavor branes:
\begin{eqnarray}
&&N_{8}^f = 2 \pi F_0 =  \frac{\pi}{2} R_-^2 \, , \qquad
N_{6}^f = \frac{1}{2\pi \kappa_{10}^2 T_6} \int_{S^2} \hat F_2 =  n N_8^f \;, \label{D8D6} \\
&&\label{D4} N^f_{4} = \frac{1}{2\pi \kappa_{10}^2 T_4} \int_{S_+^3}\int_{\rho_n}^{\rho_{n+1}}  \dd \rho\, \hat F_4  \, , \quad 
N^f_{2} = \frac{1}{2\pi \kappa_{10}^2 T_2} \left| \int_{S_+^3\times S^2}\int_{\rho_n}^{\rho_{n+1}}  \dd \rho\,\hat F_6 \right|\, .  
\end{eqnarray}
Here we have made explicit the interval on which the $\rho$ direction has to be integrated and we have not restricted ourselves to vanishing large gauge transformations. 

The first two charges in (\ref{D8D6}) correspond to the D8 and D6 flavor branes that originate on the $N_5^+$ D5-branes of the original background. Thus, our expectation is to find BPS D8 wrapped on $\{t, x_1, r_+, S^3_+, \rho, S^2\}$ and BPS D6 wrapped on $\{t, x_1, r_+, S^3_+, \rho\}$. 
However, as for the color branes, we also find that the D6 are never BPS unless $R_-= 0$ and that the D8 (anti-D8 in our conventions) are BPS only in the absence of large gauge transformations. 
This is again suggestive of a dual background where large gauge transformations are not possible. In the absence of these the D5 flavor branes give rise to just D8 flavor branes in the dual background.

The D5' branes of the original background give rise in turn to D4-branes wrapped on  $\{t, x_1, r_-, S^2\}$ and D2-branes wrapped on $\{t, x_1, r_-\}$, which turn out to be BPS only when located at $\rho=0$. In this position however both the DBI and CS actions of the D4 vanish, leaving just D2-branes as candidate flavor branes. 

\subsubsection{A possible brane intersection?}

Summarizing, we have found that there are only BPS color and flavor branes in the absence of large gauge transformations, in which case there is only one color or flavor brane in the non-Abelian T-dual background associated to each color or flavor brane of the original theory. Note that this is essentially different from previous examples in the literature (for instance \cite{Lozano:2013oma, Lozano:2014ata}) where both types of color and flavor branes were guaranteed to exist for all $n$. We argue in the conclusions that this could be explained by the absence of non-trivial 2-cycles in our particular dual geometry\footnote{In the examples in \cite{Lozano:2013oma, Lozano:2014ata} non-trivial $S^2$ were guaranteed to exist due to the presence of singularities.}.

We have shown that the D1-branes are replaced by D4-branes wrapped on 
$\{t, x_1, \rho, S^2\}$ and the D5 and D5' flavor branes are replaced by D8-branes wrapped on $\{t, x_1, r_+, S^3_+, \rho, S^2\}$ and D2-branes wrapped on $\{t, x_1, r_-\}$, respectively. This is summarized pictorially as 
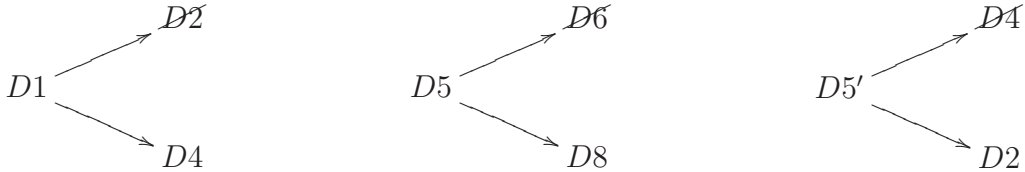
\begin{equation*}
  \xymatrixrowsep{1pc} \xymatrixcolsep{3pc} 
  \xymatrix{  & \cancel{D2} \\
     D1\ar[ur] \ar[dr] &   \\
     &   D4}
\hspace{2.5cm}
  \xymatrixrowsep{1pc} \xymatrixcolsep{3pc} 
  \xymatrix{  & \cancel{D6} \\
     D5\ar[ur] \ar[dr] &   \\
     &   D8}     
\hspace{2.5cm}
  \xymatrixrowsep{1pc} \xymatrixcolsep{3pc} 
  \xymatrix{  & \cancel{D4} \\
     D5'\ar[ur] \ar[dr] &   \\
     &   D2}       
\end{equation*}
Here we have also indicated the brane that turns out not to occur as a BPS configuration even if expected a priori from the analysis of the fluxes.

Note that precisely a $D1\rightarrow D4$, $D5\rightarrow D8$, $D5'\rightarrow D2$ map is what one would have obtained after (Abelian) T-dualizing  the D1, D5, D5' system along three directions transverse to the D1 and the D5 and longitudinal to the D5'.  
This
suggests a dual geometry coming out as the near horizon limit of the brane intersection: 
\begin{eqnarray}
&&N_8^f \,\,D8: \qquad  012345789\nonumber\\
&&N_2^f \, D2: \qquad \,  016 \nonumber \\
&&N_4 \, D4: \qquad \, \, 01789
\end{eqnarray}
In this brane intersection the $SO(4)^+\times SO(4)^-$ symmetry of the original field theory is  replaced by a $SO(4)^+\times SU(2)$ symmetry. Of this, $SU(2)_R^+\times SU(2)_R$ would correspond to the R-symmetry group of a large $\mathcal{N}=(0,4)$ field theory living at the intersection, and the remaining $SU(2)_L^+$ to a global symmetry. This is consistent with the central charge computation in subsection 4.6 and with the supersymmetry analysis in section 7 (see also the Appendix). The field theory would moreover have gauge group $U(N_4)$ and a global symmetry $SU(N_8^f)\times SU(N_2^f)$. Some field theory configurations that we present next are compatible with this brane realization.

\subsection{Instantons}

A very similar calculation to the one in subsection 3.2 shows that the D4 color branes can be realized as instantons in the D8 flavor branes. In this case
\begin{equation}
S^{D8}_{\rm fluc}=-\int\frac{1}{g_{D8}^2} F^2_{\mu\nu} \qquad {\rm with} \qquad \frac{1}{g_{D8}^2}=\frac{L^2 r_+^2 r_-^2 \rho^2}{(2\pi)^6}
\end{equation}
and the DBI action of the D4-branes satisfies
\begin{equation}
S_{DBI}^{D4}=-\int\frac{16\pi^2}{g_{D8}^2}\, ,
\end{equation}
as expected for an instantonic brane.

\subsection{Baryon vertices and t'Hooft monopoles}

The original D7-brane baryon vertex configuration is mapped after the duality into a D4-brane wrapped on $S^3_+\times S^1$ and a D6-brane wrapped on $S^3_+\times S^1 \times S^2$. The second one however has vanishing tadpole charge in the absence of large gauge transformations, given that
\begin{align}
 S_{CS}^{D6}& = 2\pi\, T_6 \int (C_5-B_2\wedge C_3 )\wedge F   
 =  -2\pi\, T_6 \int_{S^3_+\times S^1 \times S^2}\! \!\!\! \hat F_6 \int \dd t A_t \nn \\
 &= -n N_4 \int \dd t A_t \;.
\end{align}
For the $D4$ wrapped on $S^3_+\times S^1$ we find
\begin{equation}
 S^{D4}_{CS}=  -2\pi \,T_4 \int_{S^3_+\times S^1}\! \!\!\! \hat F_4 \int \dd t A_t= -N_4 \int \dd t A_t \;.
\end{equation}
As a result, there is one candidate for baryon vertex in the non-Abelian T-dual background, realized as a D4-brane wrapped on $S^3_+\times S^1$.

Similarly, in the original background we had $D3^\pm$-branes wrapped on $S^3_\pm$ t'Hooft monopoles whose tadpole charges were given by the ranks of the 
flavor groups. The $D3^+$ is mapped after the duality into a D4 wrapped on $\{S^3_+, \rho\}$ and a D6 wrapped on $\{S^3_+, \rho, S^2\}$ with tadpole charges
\begin{equation}
 S^{D4}_{CS}=  -2\pi \,T_4 \int_{S^3_+}\int_{\rho_n}^{\rho_{n+1}} \dd \rho \hat F_4 \int \dd t A_t= N_4^f \int \dd t A_t \, ,
\end{equation}
and
\begin{equation}
 S_{CS}^{D6}
 =  -2\pi T_6 \int_{S^3_+\times S^2}\int_{\rho_n}^{\rho_{n+1}}\dd \rho \hat F_6\int \dd t A_t 
 = - {N}_2^f \int \dd t A_t\;.
\end{equation}
Given that $N_4^f$ is not associated to a BPS D4-brane in the absence of large gauge transformations it is sensible to also not associate to it a 't Hooft monopole configuration. The D6-brane thus remains as the candidate 't Hooft monopole, with tadpole charge given by the charge of the D2 flavor brane. 

The $D3^-$ 't Hooft monopole of the original background is in turn mapped into a D0-brane and a D2-brane wrapped on the $S^2$. We indeed find that these branes have tadpoles with charges
\begin{equation}
 S^{D0}_{CS}=  -2\pi \,T_0 \,m \int \dd t A_t= - N_8  \int \dd t A_t \; ,
\end{equation}
and
\begin{equation}
 S^{D2}_{CS}=  -2\pi \,T_2 \int_{S^2} \hat F_2 \int \dd t A_t= - n N_8  \int \dd t A_t \;.
\end{equation}
Clearly the second brane does not carry any tadpole charge in the absence of large gauge transformations. Thus, only the D0-brane remains as candidate 't Hooft monopole, with tadpole charge given by the charge of the D8 flavor brane.

Consistently with our previous results we find two 't Hooft monopole configurations in the dual background whose tadpole charges are given by the charges of the two D2 and D8 dual flavor branes.

\subsection{Central charge}

Finally in this section we compute the central charge of the dual supergravity solution. We show that as in the original theory it is possible to define two R-symmetry currents from which
\begin{equation}
\label{sevrin}
c=2\frac{k^+ k^-}{k^++k^-}
\end{equation}
 as in  \cite{Sevrin:1988ew}.
We take the general expressions in \cite{Klebanov:2007ws}, to which the reader is referred for more details. 

Rewriting the original IIB metric as
\begin{equation}
 \dd s^2_{str}= \alpha(r) \beta(r) \dd r^2+ \alpha(r) \dd x_{1,1}^2+g_{ij}\dd y^i\dd y^j,
\end{equation}
we read off
\begin{equation}
\alpha  = L^2 r^2,~~~\beta =\frac{1}{r^4}\, .
\end{equation}
Substituting these in the expressions for the internal volume\footnote{We have generalized these as in \cite{Lozano:2014ata} to account for the $y$-dependent dilaton in the dual background.} and $r$-dependent quantity $\kappa$ we obtain
 \begin{align} \label{eq:Vint}
  V_{int} &= \int \dd^7 \!y\, e^{-2\Phi} \,\sqrt{\det (g_{ij})} = 4\pi^4 R_+^3R_-^3 \delta x \\ 
 \kappa &= V_{int}^2\, \alpha(r)=  V_{int}^2 L r    . \nn
 \end{align}
The central charge of the original theory can then be computed as
\begin{equation}
\label{prop}
c\sim \beta^{d/2}\kappa^{3d/2}(\kappa^\prime)^{-d}
\end{equation}
where $d=1$ in our case and $\kappa^\prime\equiv d\kappa/dr$, to obtain
\begin{equation}
\label{coriginal}
c = \frac{1}{(2\pi)^2} L R_+^3 R_-^3\, \delta x = 2 L^2 N_1 = 2 N_1 \frac{N_5^+ N_5^-}{N_5^+ + N_5^-} \, ,
\end{equation}
where we have substituted $\delta x$ from (\ref{eq:N1orig}), $L^2=N_5^+ N_5^-/(N_5^++N_5^-)$, and have fixed the normalization factor in (\ref{prop}) to agree with the central charge computed in  \cite{Sevrin:1988ew}, with $k^\pm = N_1 N_5^\pm$.

Similarly for the non-Abelian T-dual solution we find
\begin{equation} \label{eq:VintDual}
\tilde V_{int} = \int \dd^7 \!y \,e^{-2\Phi}\sqrt{\det (\tilde g_{ij})} = \frac{1}{3} \pi^6 R_+^3R_-^3 \delta x \;, 
\end{equation}
from where, taking the same normalization factor as in (\ref{coriginal}),
\begin{equation}
{\tilde c}=\frac{1}{48}  L R_+^3 R_-^3\, \delta x=\frac{\pi}{3}L^2 N_4=2N_4\frac{N_2^f N_8^f}{3N_2^f+N_8^f}\, .
\end{equation}
Note that it is not possible to bring the dual central charge into the form (\ref{sevrin}) unless we change the normalization factor. Indeed, the change in the internal volume produced by the non-Abelian T-duality transformation translates generically into central charges differing by constant factors
(see for instance \cite{Lozano:2013oma, Itsios:2013wd}). Still, up to this normalization factor, the central charge is of the form 
(\ref{sevrin}), with two levels that depend differently on the products of color and flavor charges. We denote these by $k^+=3 N_4 N_2^f$, $k^-=N_4 N_8^f$. Note that consistently with the form of the dual geometry, the $+\leftrightarrow -$ symmetry of the original background has now disappeared. It would be interesting to understand the field theory origin of the values for the two  levels that we obtain. The central charge is thus compatible with a large $\mathcal{N}=(0,4)$ superconformal theory dual to our solution. 

\section{Example in new class of $AdS_3 \times S^2$ geometries in 11D} 
\label{sec:masslessIIA}

In this section, following section \ref{sec:SU2structure}, we manipulate the massive IIA solution of the previous section by performing two Abelian T-dualities, in the process rendering it as a solution to massless IIA supergravity.  We will then be in a position to uplift the solution to 11D supergravity. As we detail in section \ref{sec:SUSY}, while not entirely obvious, there are indeed \textit{two} manifest global $U(1)$ isometries, namely the overall transverse $x$-direction and the remaining Hopf-fibre, which becomes a global symmetry after the initial T-duality. 

Performing the T-duality on the $x$-direction, the NS sector is unchanged, while the T-dual RR sector becomes
\bea
F_1 &=& \frac{R_-^2}{4} \dd x, \cr
F_3 &=& \frac{4 R_-^2 \, \rho^3}{16 \rho^2 + R_-^4 } \sin \chi \dd \chi \wedge \dd \xi\wedge \dd x - \frac{R_-^3}{4 R_+ L }  [ L^4 \textrm{Vol}(AdS_3) + R_+^4 \textrm{Vol}(S^3_+) ], \cr
F_5 &=&  [ 2 L^2 \textrm{Vol}(AdS_3) + 2 R_+^2 \textrm{Vol}(S^3_+) ] \wedge \rho \, \dd \rho \wedge \dd x \cr &-& \frac{4 R_-^3\, \rho^3}{ L \, R_+( 16 \rho^2 + R_-^4)}  [ L^4 \textrm{Vol}(AdS_3) + R_+^4 \textrm{Vol}(S^3_+) ] \sin \chi \dd \chi \wedge \dd \xi. 
\eea
We can further T-dualise on the Hopf-fibre direction, which we parametrise through the coordinate $\psi$, to get the massless IIA solution: 
\bea
\dd \hat{s}^2 &=& L^2 \dd s^2 (AdS_3) + \frac{R_+^2}{4} ( \dd \theta^2 + \sin^2 \theta \dd \phi^2) + \dd x^2 + \frac{4}{R_+^2} \dd \psi^2 \cr && \phantom{xxxxxxxxxxxxxxxxxx}+  \frac{4}{R_-^2} \dd \rho^2 + \frac{4 R_-^2 \, \rho^2}{16 \rho^2 + R_-^4} (\dd \chi^2 + \sin^2 \chi \dd \xi^2 ), \cr
\hat{B} &=& \frac{16 \rho^3}{16 \rho^2 + R_-^4} \sin \chi \dd \chi \wedge \dd \xi + \cos \theta \dd \phi \wedge \dd \psi, \cr
e^{-2 \hat{\Phi}} &=& \frac{R_-^2\, R_+^2}{256} ( 16 \rho^2 + R_-^4), \cr
F_2 &=& -\frac{R_-^2}{4} \dd x \wedge \dd \psi - \frac{R_-^3\, R_+^3}{32 L}  \sin \theta \dd \theta \wedge \dd \phi , \cr
F_4 &=&  -\frac{4 R_-^2 \, \rho^3}{16 \rho^2 + R_-^4 } \sin \chi \dd \chi \wedge \dd \xi\wedge \dd x \wedge \dd \psi +\frac{R_-^3 L^3}{4 R_+}  \textrm{vol} (AdS_3) \wedge \dd \psi, \cr
&+& \frac{\rho \,R_+^2}{4} \sin \theta \dd \theta \wedge \dd \phi \wedge \dd \rho \wedge \dd x - \frac{R_-^3 \, R_+^3 \, \rho^3}{2 L (16 \rho^2 + R_-^4)} \sin \theta \dd \theta \wedge \dd \phi \wedge \sin \chi \dd \chi \wedge \dd \xi. \nonumber
\eea

Uplifting to 11D, we get: 
\bea
\dd s^2_{11} &=&  e^{2 \lambda} \biggl[  L^2 \dd s^2 (AdS_3) + e^{2 A} (\dd \chi^2 + \sin^2 \chi \dd \xi^2 ) + \dd s^2_6 \biggr], \cr
G_4 &=& -\frac{4 R_-^2 \, \rho^3}{16 \rho^2 + R_-^4 } \sin \chi \dd \chi \wedge \dd \xi\wedge \dd x \wedge \dd \psi +\frac{R_-^3 L^3}{4 R_+}  \textrm{vol} (AdS_3) \wedge \dd \psi, \cr
&+& \frac{R_+^2}{4} \sin \theta \dd \theta \wedge \dd \phi \wedge \rho \dd \rho \wedge \dd x - \frac{R_-^3 \, R_+^3 \, \rho^3}{2 L (16 \rho^2 + R_-^4)} \sin \theta \dd \theta \wedge \dd \phi \wedge \sin \chi \dd \chi \wedge \dd \xi \cr
&+& \left[ \frac{16 \rho^2 ( 16 \rho^2 + 3 R_-^4)}{(16 \rho^2 + R_-^4)^2} \dd \rho \wedge \sin \chi \dd \chi \wedge \dd \xi - \sin \theta \dd \theta \wedge \dd \phi \wedge \dd \psi \right] \wedge \DD z,
\eea
where we have defined
\bea
e^{2 \lambda} &=& e^{- \frac{2}{3} \hat{\Phi}}, \quad e^{2A} = \frac{4 R_-^2 \, \rho^2}{16 \rho^2 + R_-^4}, \cr
\dd s^2_6 &=& \frac{R_+^2}{4} ( \dd \theta^2 + \sin^2 \theta \dd \phi^2) + \dd x^2 + \frac{4}{R_+^2} \dd \psi^2 +  \frac{4}{R_-^2} \dd \rho^2 +  \frac{256}{R_-^2 R_+^2 ( 16 \rho^2 + R_-^4)} \DD z^2,  \cr
\DD z &\equiv& \dd z + C_1, \cr
C_1 &=&  - \frac{R_-^2}{8} ( x \dd \psi - \psi \dd x) + \frac{R_-^3 R_+^3}{32 L} \cos \theta \dd \phi. 
\eea
One can check that the Bianchi identity and the equations of motion are satisfied. As we argue in section \ref{sec:SUSY}, this uplifted geometry is expected to be $\frac{1}{4}$-BPS. What is particularly interesting about this uplift is that the internal manifold exhibits $SU(2)$-structure, yet it is beyond the scope of the ansatz in \cite{Kim:2007hv}, since $\mathcal{A}$ and $\mathcal{G}$ in (\ref{gen_ansatz}) are clearly non-zero. This opens up the possibility that we can read off the relation between the 6D Killing spinors appearing in the more general classification \cite{Colgain:2010wb}, feed them into supersymmetry conditions and identify a more general class of supersymmetric $AdS_3 \times S^2$ solutions in 11D supergravity with $SU(2)$-structure manifolds. One can then use the supersymmetry conditions to find further explicit solutions, some of which may be, in contrast to non-Abelian T-duals, compact. We hope to report on this in future work \cite{work_in_progress}. 

\section{A new IIB $AdS_3\times S^2\times S^2$ solution}
\label{sec:SU(2)x2}

In this section we dualize once more the $AdS_3\times S^3\times S^2$ solution of section 4 with respect to the $SU(2)_L^+$ acting on the $S^3_+$. We show that this dualization produces a new $AdS_3$ solution, this time in Type IIB. As we discuss in section \ref{sec:SUSY}, and further in appendix A, the new solution we generate will be $\frac{1}{4}$-BPS and still preserve $\mathcal{N} = (0,4)$ supersymmetry in 2D \footnote{The only subtlety here would appear to be the correct identification of the global $SU(2)$ with respect to which one T-dualises.}.

The new background is given by
\begin{align} \label{eq:dual_metric_2}
 ds^2_{IIB}&= L^2 \dd s^2 (AdS_3) +\dd x^2 
  + \frac{4}{R_+^2}\bigg( \dd \rho_+^2+\frac{R_+^6 \rho_+^2}{64\Delta_+}\big(\dd \chi_+^2+\sin^2\chi_+ \dd \xi_+^2\big)\bigg) \nn \\
 &\phantom{=} + \frac{4}{R_-^2}\bigg( \dd \rho_-^2+\frac{R_-^6 \rho_-^2}{64\Delta_-}\big(\dd \chi_-^2+\sin^2\chi_- \dd \xi_-^2\big)\bigg) ,
\end{align}
where we have introduced $(\rho_-,\chi_-,\xi_-)$ to equal our previous $(\rho,\chi,\xi)$ after  the first dualization on $S_-^3$, and $(\rho_+,\chi_+,\xi_+)$ to denote the new coordinates arising after the second dualization on $S_+^3$. $\Delta_\pm$ are given by 
\beq \label{eq:detMpm}
\Delta_\pm= \frac{R_\pm^6+16R_\pm^2\rho_\pm^2}{64}.
\eeq
The corresponding dilaton is just
\beq  \label{eq:dilaton_2}
e^{-2\Phi}=\Delta_+ \Delta_-,
\eeq
and the new NS-NS 2-form is given by
\beq
\label{NSNS2form_2}
B_2= \frac{R_+^2 \rho_+^3}{4 \Delta_+} {\rm Vol}(S_+^2) +  \frac{R_-^2 \rho_-^3}{4 \Delta_-} {\rm Vol}(S_-^2)
\eeq
where $S_\pm^2$ are the 2-spheres parameterized by $(\chi_\pm, \xi_\pm)$, respectively. The dual RR sector is given by\footnote{Note that these are the fluxes associated to the Page charges.}
\begin{align}
\hat F_1 &= \frac{ R_-^3 R_+^3}{32 L}\dd x + \frac{1}{4}R_-^2\rho_+ \dd \rho_+ - \frac{1}{4}R_+^2\rho_- \dd \rho_-, \nn\\[4 mm]
\hat F_3 & =\frac{1}{4} R_+^2\rho_-^2 \,\dd \rho_- \wedge {\rm Vol}(S_-^2)   -\frac{1}{4}R_-^2\rho_+^2 \,\dd \rho_+ \wedge {\rm Vol}(S_+^2), \nn\\[4 mm]
\hat F_5 &= 2 L^2 \rho_-\rho_+ \, {\rm Vol}(AdS_3)\wedge \dd \rho_-\wedge \dd \rho_+  - \frac{L^3}{4} {\rm Vol}(AdS_3)\wedge  \dd x \wedge \left(  \frac{R_+^3}{ R_-}\rho_-\,\dd \rho_-  + \frac{R_-^3}{R_+}\rho_+\,\dd \rho_+ \right), \nn\\[4 mm]
\hat F_7 &= \frac{L^3}{4} {\rm Vol}(AdS_3)\wedge \dd x\wedge \Big( \frac{R_+^3}{R_-}\rho_-^2 \dd \rho_-\wedge {\rm Vol}(S_-^2) + \frac{R_-^3}{R_+}\rho_+^2 \dd \rho_+\wedge {\rm Vol}(S_+^2) \Big), \nn\\[4 mm]
&\phantom{=} - 2L^2 \rho_-\rho_+ {\rm Vol}(AdS_3)\wedge \dd \rho_-\wedge \dd \rho_+\wedge \left( \rho_- {\rm Vol}(S_-^2) + \rho_+ {\rm Vol}(S_+^2) \right),  \nn\\[4 mm]
\hat F_9 &= 2 L^2 \rho_-^2\rho_+^2 {\rm Vol}(AdS_3)\wedge \dd \rho_-\wedge{\rm Vol}(S_-^2)\wedge \dd \rho_+\wedge{\rm Vol}(S_+^2).
\end{align}

This solution satisfies the IIB equations of motion and preserves eight supersymmetries. As our previous massive $AdS_3$ solution, it is perfectly regular, with the range of the new $\mathbb{R}^+$ direction, $\rho_+$, also to be determined. As we did after the first dualization, we link the running of both non-compact directions $\rho_\pm$ to large gauge transformations in this background. The ranges of these coordinates must then be divided in $[\rho_{\pm(n_\pm)}, \rho_{\pm (n_\pm +1)}]$ intervals in which large gauge transformations with $n_\pm$ parameters on the non-trivial $S^2_\pm$ cycles ensure that $B_2$ lies in the fundamental region. 

The field theory analysis that can be made from this supergravity solution follows very closely the one we made for the previous massive $AdS_3$ solution, so we will omit the details. As in that case each of the brane configurations that we described in section 2 is mapped to a single brane configuration in the dual for $n_\pm=0$, and no dual configurations exist otherwise unless $R_-=R_+=0$. For $n_\pm=0$ we find the brane configurations:

\begin{itemize}
\item Color branes: D7 on $\{t, x_1, \rho_-, S^2_-, \rho_+, S^2_+\}$
\item Flavor branes: D5 on $\{t, x_1, r_-, \rho_+, S^2_+\}$ (at $\rho_-=0$)\\
\hspace*{2.6cm} D5' on $\{t, x_1, r_+, \rho_-, S^2_-\}$  (at $\rho_+=0$) 
\end{itemize}
\noindent This can be summarized pictorially as 

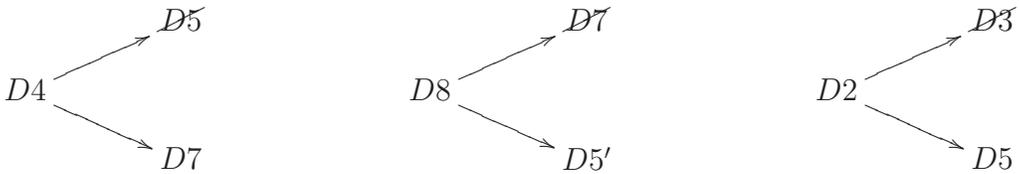
\begin{equation*}
  \xymatrixrowsep{1pc} \xymatrixcolsep{3pc} 
  \xymatrix{  & \cancel{D5} \\
     D4\ar[ur] \ar[dr] &   \\
     &   D7}
\hspace{2.5cm}
  \xymatrixrowsep{1pc} \xymatrixcolsep{3pc} 
  \xymatrix{  & \cancel{D7} \\
     D8\ar[ur] \ar[dr] &   \\
     &   D5'}     
\hspace{2.5cm}
  \xymatrixrowsep{1pc} \xymatrixcolsep{3pc} 
  \xymatrix{  & \cancel{D3} \\
     D2\ar[ur] \ar[dr] &   \\
     &   D5}       
\end{equation*}
where we have crossed out the branes not occurring as BPS configurations but expected a priori from the analysis of the fluxes. 
The charges of the surviving BPS D7, D5 and D5' are:
\begin{align}
N_7 &= +\frac{1}{2 \kappa_{10}^2 T_7} \int_{S^1} \hat F_1= \frac{R_+^3 R_-^3}{32 L} \delta x \\ 
N^+_5 &= - \frac{1}{2 \kappa_{10}^2 T_5} \int_{S_-^2} \int_0^{\rho_{-(1)}} \dd \rho_-\, \hat F_3  \\ 
N^-_{5} &= + \frac{1}{2 \kappa_{10}^2 T_5} \int_{S_+^2} \int _0^{\rho_{+(1)}} \dd \rho_+\, \hat F_3   
\end{align}
where once again $\delta x$ is the hand-set length of the $x$-direction. $\rho_{\pm (1)}$ satisfy
$16\rho_{\pm (1)}^3/(R_\pm ^4+16\rho_{\pm (1)}^2)=\pi$.
Hence, a candidate brane intersection is:
\begin{align}
\label{braneinter}
N_5^+\,\,D5: &\qquad  013456\nonumber\\[1.5 mm]
N_{5}^- \, D5':& \qquad  012789 \nonumber \\[1.5 mm]
N_7 \, D7: &\qquad  01345789
\end{align}
which realizes the $SU(2)^+\times SU(2)^-$ symmetries of the background. As shown in section 7 (see also the Appendix) these correspond to R-symmetries in the dual theory. Thus, the dual field theory is still  a large $\mathcal{N}=(0,4)$ SCFT.
The field theory living at the intersection would have gauge group $U(N_7)$ and a global symmetry $SU(N_5^+)\times SU(N_{5}^-)$.

\noindent Consistently with this picture we also have:

\begin{itemize}
\item Baryon vertices: D1 on $\{t, S^1\}$ with tadpole charge $N_7$
\item 't Hooft monopoles: ${\rm D3}^\pm$ on $\{t, \rho_\pm, S^2_\pm\}$  with tadpole charge $N_5^\mp$
\item Central charge: 
\begin{equation}
c= \frac{2}{3} N_7 \frac{N_5^+ N_5^-}{N_5^+ + N_5^-}
\end{equation}
\end{itemize}

This form for the central charge agrees with a large $\mathcal{N}=(0,4)$ dual CFT with affine $SU(2)^\pm$ current algebras at levels $k^\pm=N_7N_5^\pm$, even if with a different overall factor compared to 
\cite{Sevrin:1988ew}. This is consistent with the supersymmetry analysis.
Together with the analysis of brane configurations this suggests a dual field theory in which D7 branes substitute the D1-branes of the original field theory dual to the $AdS_3 \times S^3_+\times S^3_-\times S^1$ solution. In this theory the global $SU(2)_+^L\times SU(2)_-^L$ symmetries have disappeared. It would be interesting to see if one can indeed derive these properties from the brane intersection given by (\ref{braneinter}).

\section{Comments on supersymmetry} 
\label{sec:SUSY}
In this section we comment on the number of supersymmetries the various solutions to 10D Type II supergravity preserve. To make the text self-contained, we start by recalling our supersymmetry conventions \cite{Itsios:2012dc, Hassan:1999bv}. The fermionic supersymmetry variations for Type IIA and Type IIB supergravity are respectively 
\bea
\delta \lambda &=& \frac{1}{2} \slashed{\partial} \Phi \eta - \frac{1}{24} \slashed{H}_3 \sigma_3 \eta + \frac{1}{8} e^{\Phi} \left[ 5 m \sigma_1 + \frac{3}{2} \slashed{F}_2 i \sigma_2 + \frac{1}{24} \slashed{F}_4 \sigma_1 \right] \eta, \cr
\delta \Psi_{\mu} &=& \nabla_{\mu} \eta - \frac{1}{8} H_{3 \, \mu \nu \rho} \Gamma^{\nu \rho} \sigma_3 + \frac{1}{8} e^{\Phi} \left[ m \sigma_1 + \frac{1}{2} \slashed{F}_2 i \sigma_2 + \frac{1}{24} \slashed{F}_4 \sigma_1 \right] \Gamma_{\mu} \eta, 
\eea
and 
\bea
\delta \lambda &=& \frac{1}{2} \slashed{\partial} \Phi \eta - \frac{1}{24} \slashed{H}_3 \sigma_3 \eta + \frac{1}{2} e^{\Phi} \left[   \slashed{F}_1 i \sigma_2 + \frac{1}{12} \slashed{F}_3 \sigma_1 \right] \eta, \cr
\delta \Psi_{\mu} &=& \nabla_{\mu} \eta - \frac{1}{8} H_{3 \, \mu \nu \rho} \Gamma^{\nu \rho} \sigma_3 - \frac{1}{8} e^{\Phi} \left[ \slashed{F}_1 i \sigma_2 + \frac{1}{6} \slashed{F}_3 \sigma_1 + \frac{1}{240} \slashed{F}_5 i \sigma_2 \right] \Gamma_{\mu} \eta, 
\eea
where $\lambda$ denotes the dilatinos, $\Psi_{\mu}$ the gravitinos and $\eta$ is a Majorana-Weyl spinor 
\be
\eta = \left( \begin{array}{c} \epsilon_+ \\ \epsilon_- \end{array} \right). 
\ee
 
The supersymmetry preserved by the non-Abelian T-dual of $AdS_3 \times S^3 \times CY_2$ is well-documented \cite{Sfetsos:2010uq, Itsios:2012dc} and analysis leads to the conclusion that half the supersymmetry is broken in the transformation. Therefore, for the geometries exhibited in section \ref{sec:SU2structure}, all solutions preserve eight supersymmetries, or $\mathcal{N} = (0,4)$ supersymmetry in 2D. We have noted that the 11D uplift fits into the classification of \cite{Kim:2007hv} and further demonstrated that supersymmetry is not enhanced beyond $\frac{1}{4}$-BPS in 11D, thus providing the first concrete example in the class of \cite{Kim:2007hv}. 

For the geometry $AdS_3 \times S^3 \times S^3 \times S^1$, supersymmetry breaking is not a foregone conclusion. To see why this may be the case, we recall that the geometry  $AdS_3 \times S^3 \times S^3 \times S^1$ possesses an $SU(2) \times SU(2)$ R-symmetry, yet is manifestly $SO(4) \times SO(4)$-invariant. Therefore, it could be expected that a judicious choice of the T-duality $SU(2)$ factor would result in a geometry preserving the same amount of supersymmetry as the original solution. This intuition is based on ref. \cite{Lozano:2012au}, where T-duality with respect to a global $SU(2)$ isometry generated a surprising new supersymmetric $AdS_6$ solution to IIB supergravity. 

Here we correct statements in the literature \footnote{It was initially reported in ref. \cite{Kelekci:2014ima} that  an application of non-Abelian T-duality to an $SU(2)$ factor in one of the $SO(4)$ isometries resulted in a T-dual preserving sixteen supersymmetries. The analysis of ref. \cite{Kelekci:2014ima} failed to take account of an additional condition, which breaks supersymmetry to eight.} and show that picking out a left or right-acting $SU(2)$ isometry from one of the three-spheres leads to broken supersymmetry in an analogous fashion to $AdS_3 \times S^3 \times CY_2$ non-Abelian T-duals. For completeness, we do this in two ways, uncovering a consistent picture. 

Firstly, and most easily, we can import the findings of ref. \cite{Itsios:2012dc}. We recall for spacetimes with $SO(4)$ isometry - with generalisations to $SU(2)$ isometry \cite{Kelekci:2014ima} - that supersymmetry breaking is encoded in a single condition,  namely (3.11) of ref. \cite{Itsios:2012dc},
\be
\label{original_cond}
\biggl[ - \frac{1}{2 R_-} \Gamma^{\chi \xi} \sigma_3  - \frac{1}{4 R_- } \Gamma^{\chi \xi  \rho} i \sigma_2- \frac{1}{4} \left(\frac{1}{L} \Gamma^{012} + \frac{1}{R_+} \Gamma^{678} \right) \sigma_1 \biggr] \tilde{\eta} = 0, 
\ee
where, assuming we T-dualise from the IIB form for the geometry, $\tilde{\eta}$ is related by a factor to the Killing spinor of IIA supergravity $\eta$, 
\be
\label{spinor_transform}
\tilde{\eta} \equiv e^{-X} \eta = \exp \left( \frac{1}{2} \tan^{-1} \left( \frac{R^2_-}{4 \rho} \right) \Gamma^{\chi \xi} \sigma_3 \right) \eta.
\ee

Using (\ref{Lerres}), we can rewrite these conditions as: 
\bea
\label{eplus}
\biggl[ -\frac{1}{R_-} \Gamma^{\chi \xi } + \frac{1}{L} \Gamma^{012 \rho} + \frac{1}{R_+} \Gamma^{678 \rho} \biggr] \tilde{\epsilon}_+ &=& 0, \cr 
\label{eminus}
\biggl[ \frac{1}{R_-} \Gamma^{\rho \chi \xi} - \frac{1}{L} \Gamma^{012} - \frac{1}{R_+} \Gamma^{678} \biggr] \tilde{\epsilon}_- &=& 0, 
\eea
and further using (3.14) of ref. \cite{Itsios:2012dc}, which in this case reads, 
\be
\tilde{\epsilon}_+ = \Gamma^{\rho} \epsilon_+, \quad \tilde{\epsilon}_- = - \epsilon_-, \quad \Gamma^{\rho \chi \xi} = - \Gamma^{345},
\ee
we recover what turns out to be the original projection condition of the IIB geometry (\ref{original})
\be
\label{orig_proj}
\left[ \frac{1}{L} \Gamma^{012} + \frac{1}{R_+} \Gamma^{345} + \frac{1}{R_-} \Gamma^{678} \right] {\eta} =0. 
\ee 
We observe that squaring this expression, we recover (\ref{Lerres}). We note also that in the process of redefining the spinors, the chirality of $\tilde{\epsilon}_{+}$ is flipped so that it now corresponds to a Killing spinor of Type IIB supergravity. 
On its own, this projection condition would suggest the background is $\frac{1}{2}$-BPS, however we also find that the following identification is also implied
\be
\label{identity}
\epsilon_+ = \epsilon_-. 
\ee
This constitutes an additional condition, which breaks supersymmetry to $\frac{1}{4}$-BPS, or eight supersymmetries. 


To develop a better understanding of what has just happened, it is also useful to explicitly work out the Killing spinors for the original solution (\ref{original}). Following a calculation similar to ref. \cite{Gauntlett:1998kc}, except translated into our conventions, and making use of the projection condition (\ref{orig_proj}), which falls out from the analysis, we can determine the precise form of the Killing spinors in their original IIB setting:  
\bea
\epsilon_+ &=& \left[  r^{\frac{1}{2}} + r^{- \frac{1}{2}} ( t \Gamma_{0}^{~2} + x \Gamma_{1}^{~2}) \right] ( \alpha_1 +  \Omega \beta_1) + r^{-\frac{1}{2}} ( \Omega \alpha_2 + \beta_2), \cr
\epsilon_- &=& \left[  r^{\frac{1}{2}} + r^{- \frac{1}{2}} ( t \Gamma_{0}^{~2} + x \Gamma_{1}^{~2}) \right] ( \alpha_1 -  \Omega \beta_1) - r^{-\frac{1}{2}} ( \Omega \alpha_2 - \beta_2), 
\eea 
where we have defined the constant spinors,  $\Gamma^{01} \alpha_i = \alpha_i$, $\Gamma^{01} \beta_i = - \beta_i$, and the matrix, 
\be
\Omega = e^{-\frac{1}{2} \psi_1 \Gamma^{34}} e^{- \frac{1}{2} \theta_1 \Gamma^{53}} e^{-\frac{1}{2} \phi_1 \Gamma^{34}} e^{-\frac{1}{2} \psi_2 \Gamma^{67}} e^{- \frac{1}{2} \theta_2 \Gamma^{86}} e^{-\frac{1}{2} \phi_2 \Gamma^{67}}, 
\ee
where the angular dependence follows for the explicit form of left-invariant one-forms $\tau_{\alpha}$, which satisfy $\dd \tau_{\alpha} = \frac{1}{2} \epsilon_{\alpha \beta \gamma} \tau^{\beta} \wedge \tau^{\gamma}$. The existence of Poincar\'e supersymmetries of both chirality with respect to $\Gamma^{01}$ indicates that supersymmetry is $\mathcal{N} = (4,4)$ in 2D.

We now can appreciate that the identification (\ref{identity}) is a by-product of the fact that all Killing spinors with angular dependence get projected out under the non-Abelian T-duality. It is worth noting that a single Hopf-fibre T-duality also results in the same supersymmetry breaking, although one can consider a linear combination of the Hopf-fibres, which preserves additional supersymmetries \cite{Donos:2014eua} \footnote{That the T-dual geometry in this special case must preserve twelve supersymmetries, and not the generic eight, can be most easily seen by resorting to the Kosmann spinorial-Lie-derivative \cite{Kosmann}. One can then use the powerful result in ref. \cite{Kelekci:2014ima} that the supersymmetries uncharged under the T-duality direction are preserved.}. 

This final observation that angular dependence gets projected out presents us with a small puzzle. Namely, how can the loss of angular dependence be reconciled with $\mathcal{N} = (0,4)$ supersymmetry, which requires, at a very least, the geometric realisation of an associated $SU(2)$ R-symmetry? To answer this question, we need to recall that an $SU(2)$ transformation on a round three-sphere results in a residual $S^2$ factor in the metric. This then is  one candidate $SU(2)$ R-symmetry. As we shall appreciate later, the Killing spinors of the non-Abelian T-dual also have dependence on $SU(2)_{R}$ of the remaining three-sphere. This suggests the presence of large $\mathcal{N} = (0,4)$ supersymmetry where the corresponding isometry group is $D(2, 1 | \gamma) \times SL(2, \mathbb{R}) \times SU(2)$, which as we explain in the appendix, is analogous to the Abelian T-dual, i. e. the geometry $AdS_3 \times S^3 \times S^2 \times T^2$. 

In a bid to make this work self-contained, we now explicitly check that the residual $S^2$ becomes the $SU(2)$ R-symmetry, that the remaining $SO(4)$  has an $SU(2)_{L}$ global symmetry and that supersymmetry is indeed $\mathcal{N} = (0,4)$, as claimed. To do so, we solve the Killing spinor equations for IIA supergravity in the T-dual geometry. 

We begin by introducing a frame for the remaining three-sphere, 
\be
\dd s^2(S^3_+) = \frac{1}{4} [ ( \dd \psi + \cos \theta \dd \phi)^2 + \dd \theta^2 + \sin^2 \theta \dd \phi^2]. 
\ee 
We next introduce the natural dreibein, $e^{6} = \frac{R_+}{2} ( \dd \psi + \cos \theta \dd \phi), e^{7} = \frac{R_+}{2} \dd \theta, e^{8} = \frac{R_+}{2} \sin \theta \dd \phi$ and reverse the overall sign of the RR sector relative to (\ref{dual_RR}), so we can import results from ref. \cite{Itsios:2012dc}, where expressions are given in terms of spherical coordinates, which are best suited to the current example. We note that $H_3 = \dd B_2$ has no legs along the $\psi$-direction, so the gravitino variation in this direction simply reads: 
\bea
\label{grav6}
e^{-X} \delta \Psi_6 &=&  \frac{2}{R_+} \partial_{\psi} \tilde{\eta} + \frac{1}{2\, R_+} \Gamma^{78} \tilde{\eta} 
+ \frac{e^{-2X}}{R_- \sqrt{16 \rho^2 + R_-^4}} \biggl[ - \frac{R_-^2}{4} \sigma_1- \rho \Gamma^{\chi \xi} i \sigma_2 \cr 
&-& \rho \left( \frac{R_-}{L} \Gamma^{012 \rho} + \frac{R_-}{R_+} \Gamma^{678 \rho} \right) \sigma_1 - \frac{R_-^3}{4} \left( \frac{1}{R_+} \Gamma^{0129} + \frac{1}{L} \Gamma^{6789}  \right) \sigma_1 \biggr]  \Gamma_6 \tilde{\eta}, 
\eea
where we have multiplied by the matrix $e^{-X}$ and redefined the original Killing spinor as in (\ref{spinor_transform}). The Killing spinor $\tilde{\eta}$ is a IIA spinor satisfying the projection conditions 
\bea
\label{IIA_proj}
 \left( \frac{R_-}{L} \Gamma^{012 \chi \xi} i \sigma_2 + \frac{R_-}{R_+} \Gamma^{678 \chi \xi} i \sigma_2 \right) \tilde{\eta}  &=& \tilde{\eta}, \cr
\Gamma^{\rho} \sigma_1 \tilde{\eta}  &=& - \tilde{\eta}, 
\eea
hopefully making it obvious, through the appearance of two projection conditions,  that the number of preserved supersymmetries is eight. 

Bearing in mind that $\tilde{\eta}$ is comprised of Majorana-Weyl spinors of opposite chirality, we can dualise gamma matrices as follows
\be
\label{dual_gamma}
\Gamma^{012 x} \sigma_1 \tilde{\eta} = \Gamma^{678 \rho \chi \xi} i \sigma_2 \tilde{\eta}, \quad \Gamma^{678 x} \sigma_1 \tilde{\eta} = \Gamma^{012 \rho \chi \xi} i \sigma_2 \tilde{\eta}. 
\ee
Then using the above expressions, one can rewrite (\ref{grav6}) as 
\bea
e^{-X} \delta \Psi_6 &=&  \frac{2}{R_+} \partial_{\psi} \tilde{\eta} + \frac{1}{2\, R_+} \Gamma^{78} \tilde{\eta} 
+ \frac{e^{-2X}}{R_+ \sqrt{16 \rho^2 + R_-^4}} \Gamma^{78} \biggl[ - 2 \rho + \frac{R_-^2}{2} \Gamma^{\chi \xi} \sigma_3 \biggr]  \tilde{\eta}. 
\eea
Finally, we insert the expression for $e^{-2 X}$, 
\be
e^{-2X} = \frac{1}{\sqrt{16 \rho^2 + R_-^4}} ( 4 \rho + R_-^2 \Gamma^{\chi \xi} \sigma_3 ),  
\ee
to reach the conclusion that $\partial_{\psi} \eta = \partial_{\psi} ( e^{X} \tilde{\eta} ) = 0$, so after an $SU(2)$ transformation the Killing spinors are independent of the Hopf-fibre, meaning that we can Abelian T-dualise later with respect to this direction. Similar calculations for the $\theta$ and $\phi$-directions show that the Killing spinors also do not depend on these. We therefore see in an explicit fashion that the second three-sphere is now comprised of a global $SU(2)_L$ symmetry, yet with the Killing spinors still dependent on $SU(2)_R$. In analogy with the Abelian case, we have an $D(2|1, \gamma) \times SL(2, \mathbb{R}) \times SU(2)$ symmetry algebra. 

To extract the R-symmetry dependence on the residual $S^2$, we consider $e^{-X} \delta \Psi_{\alpha}$, where $\alpha \in \{ \chi, \xi \}$. We find 
\bea
e^{-X} \delta \Psi_{\chi} &=& \frac{\sqrt{16 \rho^2 + R_-^4}}{2 R_- \rho} \partial_{\chi} \tilde{\eta} + \frac{e^{-2X}}{(16 \rho^2 + R_-^4)} \left( - \frac{R_-^5}{4 \rho}  \Gamma^{\chi} \sigma_1 + \frac{(16 \rho^2 + 3 R_-^4)}{2 R_-} \Gamma^{\xi} i \sigma_2 \right) \tilde{\eta} \cr
&+& \frac{1}{R_- \sqrt{16 \rho^2 + R_-^4}} \left( - \frac{R_-^2}{2} \Gamma^{\chi} \sigma_1 + 2 \rho \Gamma^{\xi} i \sigma_2  \right) \tilde{\eta},  \cr
&=& \frac{\sqrt{16 \rho^2 + R_-^4}}{2 R_- \rho} \left( \partial_{\chi} \tilde{\eta} + \frac{1}{2} \Gamma^{\xi} i \sigma_2 \right) \tilde{\eta}, 
\eea 
where in the second line we have expanded $e^{-2X}$. A similar calculation for $e^{-X} \delta \Psi_{\xi}$, after simplifications leads to 
\be
e^{-X} \delta \Psi_{\xi} =  \frac{\sqrt{16 \rho^2 + R_-^4}}{2 R_- \rho} \left( \frac{1}{\sin \chi} \partial_{\xi} \tilde{\eta} + \frac{1}{2} \frac{\cos \chi}{\sin \chi} \Gamma^{\xi \chi} - \frac{1}{2} \Gamma^{\chi} i \sigma_2 \right) \tilde{\eta}. 
\ee
Up to the inclusion of the Killing spinors for $AdS_3$, we can then write the explicit form for the IIA Killing spinor 
\be
\eta = e^{X} e^{- \frac{1}{2} \chi \Gamma^{\xi} i \sigma_2} e^{- \frac{1}{2} \xi \Gamma^{\chi \xi}} \tilde{\eta}_{AdS_3}
\ee 
where $\tilde{\eta}_{AdS_3}$ denotes the Killing spinors for $AdS_3$, 
\be
\nabla_{\mu} \tilde{\eta} = \frac{1}{2} \gamma_3 \Gamma_{\mu} \tilde{\eta}, 
\ee
where we have defined $\gamma_3 \equiv \Gamma^{012}$. A calculation similar to appendix A, then shows that supersymmetry is indeed $\mathcal{N} = (0,4)$. Similar calculations to above show that the dilatino variation vanishes. Again these results are all expected and follow from the analysis presented in \cite{Itsios:2012dc}, and more generally \cite{Kelekci:2014ima}. 

To go from the massive IIA solution of section \ref{sec:massiveIIA} to the massless solution in section \ref{sec:masslessIIA}, we perform two T-dualities with respect to both the overall transverse direction $x$ and the Hopf-fibre of the remaining three-sphere. As we have argued, both correspond to global $U(1)$ isometries and it is expected that supersymmetry will be preserved. As one further final check that this is indeed the case, we record some of the  gravitino variations after these two Abelian T-dualities. The gravitino variations in the $x$-direction and $\psi$-direction, notably those featuring in the T-duality, are respectively 
\be
e^{-X} \delta \Psi_{x} = \frac{1}{4 L} \Gamma^{\theta \phi \chi \xi} \left[ -\frac{L}{R_-} \Gamma^{\theta \phi x \psi} + \frac{L}{R_+} \Gamma^{\rho \chi \xi x} - 1\right] \sigma_1 \tilde{\eta}, 
\ee
and
\be
e^{-X} \delta \Psi_{\psi} = \frac{1}{2 R_+} \Gamma^{\theta \phi} \sigma_3 \left[ \Gamma^{\rho x \psi} i \sigma_2 + 1 \right] \tilde{\eta}  + \frac{1}{4 L} \Gamma^{\psi \theta \phi \chi \xi} \left[ -\frac{L}{R_-} \Gamma^{\theta \phi x \psi} + \frac{L}{R_+} \Gamma^{\rho \chi \xi x} -1\right] \sigma_1 \tilde{\eta},  
\ee
leading to good, commuting projection conditions. Furthermore, up to a redefinition in $\tilde{\epsilon}_{+}$, namely $\tilde{\epsilon}_+ \rightarrow -\Gamma^{x \psi} \tilde{\epsilon}_+$, with $\epsilon_{-}$ unchanged so it maintains its chirality, these projection conditions can be mapped back to (\ref{IIA_proj}), so we see that they are consistent. Yet again, by analogy with the Abelian T-duals discussed in the appendix, the isometry group for this geometry is expected to be $D(2|1, \gamma) \times SL(2, \mathbb{R})$.

\section{Conclusions}
Non-Abelian T-duality is a symmetry of the equations of motion of type II supergravity. This has been shown explicitly for $SO(4)$-invariant spacetimes  via dimensional reduction \cite{Itsios:2012dc}, results of which featured prominently in this current work. For spacetimes with less symmetry, e. g. the class of Bianchi IX spacetimes with $SU(2)$ isometry, partial results exist \cite{ Kelekci:2014ima, Jeong:2013jfc}, but given the number of examples explored to date, it is safe to assume that the non-Abelian T-duality procedure with RR fluxes outlined in \cite{Sfetsos:2010uq}, and generalised to larger non-Abelian groups in \cite{Lozano:2011kb}, will take Type II supergravity solutions into each other. 

We have made use of this solution-generating property in this paper to provide sample geometries for a class of $\frac{1}{4}$-BPS $AdS_3 \times S^2$ spacetimes in 11D supergravity, where the internal space is an $SU(2)$-structure manifold. Despite a number of studies asserting that the class exists \cite{Gauntlett:2006ux, MacConamhna:2006nb, Colgain:2010wb}, most notably the classification in  \cite{Kim:2007hv}, there was no explicit example known. Not only have we demonstrated that the non-Abelian T-dual of the well-known geometry $AdS_3 \times S^3 \times CY_2$ provides an example in this class, we have exhibited non-Abelian T-duals of a related geometry, $AdS_3 \times S^3 \times S^3 \times S^1$, which fall outside this class. This suggests that the general supersymmetry conditions of ref. \cite{Colgain:2010wb} can be mined further to extract a larger class of supersymmetric solutions based on $SU(2)$-structure manifolds, thus extending the \cite{Kim:2007hv} class. It may be hoped that the non-Abelian T-duals, despite being manifestly non-compact, may serve to identify compact solutions via ansatz when the full class of supersymmetric $AdS_3 \times S^2$ solutions of 11D supergravity are identified. 

On a related note, the $\frac{1}{4}$-BPS $AdS_3$ solutions we generate involve a Romans' mass. Therefore, they will serve as a test of an ongoing program of work classifying the $AdS$ solutions of massive IIA supergravity \cite{Passias:2012vp, Apruzzi:2013yva, Apruzzi:2015zna}. Furthermore, it may be interesting to consider non-Abelian T-duals of general $AdS_3 \times S^3 \times S^3 \times \Sigma_{2}$  solutions to 11D supergravity \cite{Bachas:2013vza}, where $\Sigma_2$ is a Riemann surface. 

We have discussed some properties of the field theories associated to the $AdS_3\times S^3\times S^2$ and $AdS_3\times S^2\times S^2$ backgrounds that we construct with an aim at testing the general ideas on the CFT interpretation of non-Abelian T-duals in  \cite{Lozano:2014ata} (see also \cite{Lozano:2013oma}). We have seen that as in previous examples there seems to be a doubling of charges after the transformation. In our $AdS_3$ cases however  the branes responsible for the extra charges turn out to be supersymmetric only in the absence of large gauge transformations, in which case the extra charges vanish.
The absence of large gauge transformations can be explained in turn either by the non-existence of non-trivial 2-cycles in the dual geometry at finite $\rho$, or, else, by a geometry terminating at a regular point.  As in \cite{Lozano:2013oma} the termination of the geometry at a regular point is intimately related to the depletion of the rank of one of the gauge groups. 

An important piece of information about the CFT duals to the new solutions comes from the analysis of their central charges. We have shown that as in the original theory it is possible to define two R-symmetry currents from which the central charges exhibit the expected 
$c\sim k^+k^-/(k^++k^-)$ behaviour for a large $\mathcal{N}=(0,4)$ superconformal algebra, in full agreement with the supersymmetry properties of the solutions.

\subsection*{Acknowledgements}

We would like to thank \"Ozg\"ur Kelekci, Nakwoo Kim, Andrea Prinsloo, Martin Ro\v{c}ek, Diego Rodr\'{\i}guez-G\'omez and Dimitrios Tsimpis for useful discussions. The work of Y.L., N.T.M. and J.M. has been partially supported by the COST Action MP1210 ``The String Theory Universe''. Y.L. and J.M. are partially supported by the Spanish Ministry of Economy and Competitiveness grant FPA2012-35043-C02-02. J.M. is supported by the FPI fellowship BES-2013-064815 linked to the previous project. He is grateful for the warm hospitality extended by the theoretical physics group at Milano-Bicocca U. where part of this work was done. 
N.T.M is supported by INFN and by the European Research Council under the European Union's Seventh Framework Program (FP/2007-2013) - ERC Grant Agreement n. 307286 (XD-STRING). E. \'O C is grateful to KIAS for hospitality during the final stages of this project. E. \'O C. is supported by the Marie Curie award PIOF-2012-328625 T-DUALITIES. 

\appendix 

\section{Hopf-fibre T-duality for $AdS_3 \times S^3 \times S^3 \times S^1$}
To support claims in the text concerning the isometry supergroup for non-Abelian T-duals, here we present simpler Hopf-fibre T-duals in an analogous fashion. Abelian Hopf-fibre T-duals of the related IIB geometry with small superconformal symmetry, namely $AdS_3 \times S^3 \times CY_2$, were considered in \cite{Duff:1998cr}. There it was noted that supersymmetry can be preserved completely. Here we explicitly show that this is not the case when one starts with a geometry with large superconformal symmetry. Moreover, following \cite{Duff:1998cr}, we could extend our analysis here to geometries supported by both NS and RR fields, where T-duality results not in $S^1 \times S^2$, but in (squashed) Lens spaces, $S^3/\mathbb{Z}_p$, however we focus on the simplest case with just RR fields. 

Starting from $AdS_3 \times S^3 \times S^3 \times S^1$ (\ref{original}), it is known that Abelian T-duality on a given Hopf-fibre will produce a $\frac{1}{4}$-BPS $AdS_3 \times S^3 \times S^2 \times T^2$ geometry, where the corresponding supergroup is $D(2| 1, \gamma) \times SL(2,\mathbb{R}) \times SU(2)$ \cite{Boonstra:1998yu, Gauntlett:1998kc}. Here $\gamma$ is a real parameter equating to the ratio of the radii of the three-sphere and two-sphere \footnote{Prior to T-duality, this is just the ratio of the radii of the three-spheres.}. Recalling that the bosonic subgroup of the supergroup $D(2| 1, \gamma)$ is $SL(2,\mathbb{R}) \times SU(2) \times SU(2)$, we recognise that the symmetries simply correspond to the isometries of $AdS_3 \times S^3 \times S^2$. 

Assuming we begin in Type IIB with the solution (\ref{original}), the geometry resulting from a Hopf-fibre T-duality may be written as
\begin{align}
 \label{U1_tdual}
 \dd s^2 &= L^2 \dd s^2_{AdS_3} + \frac{4}{R_+^2} \dd \psi^2+  \frac{R_+^2}{4} ( \dd \theta^2 + \sin^2 \theta \dd \phi^2) + R_-^2 \dd s^2_{S^3_-} + \dd x^2, \nn\\
 B_2 &= \cos \theta \dd \phi \wedge \dd \psi, \quad e^{\Phi} = \frac{2}{R_+}, \nn\\
 F_2 &= -\frac{R_+^2}{4} \sin \theta \dd \theta \wedge \dd \phi, \nn\\[2mm]
 F_4 &= \left[ 2 L^2 \textrm{Vol}(AdS_3) + 2 R_-^2  \textrm{Vol}(S^3_-) \right] \wedge \dd \psi.     
\end{align}
As with the original geometry, the Bianchis and the equations of motion are trivially satisfied. Plugging this solution into the dilatino variation, one can extract two commuting projection conditions, 
\bea
\left( - \frac{1}{R_+} \Gamma^{\theta \phi} i \sigma_2 + \frac{1}{L} \Gamma^{012 \psi} \sigma_1 + \frac{1}{R_-} \Gamma^{678 \psi} \sigma_1 \right) \eta &=&
(\Gamma^{\psi} \sigma_1 - 1 ) \eta = 0, 
\eea
confirming that we now have eight preserved supersymmetries versus the original sixteen. As a consistency check, we observe that squaring the first projection condition, we recover the constraint on the radii (\ref{Lerres}). Solving for the Killing spinor along the internal directions, we find
\be
\eta = e^{- \frac{1}{2} \theta \Gamma^{\phi} i \sigma_2} e^{\frac{1}{2} \phi \Gamma^{\theta \phi}} \tilde{\eta}, 
\ee
where $\tilde{\eta}$ denotes the Killing spinor for $AdS_3$. Employing the left-invariant one-forms for the inert three-sphere, we see that angular dependence drops out, so we have an SU(2)$_{L}$ global symmetry, just as we witnessed in the non-Abelian case.  As a direct consequence, the Killing spinors are independent of the Hopf-fibre and we can perform a further Abelian T-duality. It is an interesting feature of this geometry that uplifting on the M-theory circle to 11D, we recover the $AdS_3 \times S^3 \times S^3 \times T^2$ geometry in 11D, 
so that supersymmetry is restored to $\frac{1}{2}$-BPS \footnote{This is the reverse of the dimensional reduction considered in ref. \cite{Gauntlett:1998kc}. }. This uplift should be contrasted with the more trivial T-duality on the $x$-direction and uplift, which leads to the same upstairs solution. 

It is instructive to perform another Hopf-fibre T-duality, thus mirroring the combination of non-Abelian transformations in section \ref{sec:SU(2)x2}. Doing so with respect to the $\psi_2$-direction, we get 
%
%
\begin{align}
 \label{U1_U1_tdual}
 \dd s^2 &= L^2 \dd s^2_{AdS_3} + \frac{4}{R_+^2} \dd \psi_1^2+  \frac{R_+^2}{4} ( \dd \theta_1^2 + \sin^2 \theta_1 \dd \phi_1^2) + \frac{4}{R_-^2} \dd \psi_2^2 \nn\\ 
         &\phantom{xxxxxxxxxxxxxxxxxxxxxxxx} +  \frac{R_-^2}{4} ( \dd \theta_2^2 + \sin^2 \theta_2 \dd \phi_2^2) + \dd x^2, \nn\\
 B_2 &= \cos \theta_1 \dd \phi_1 \wedge \dd \psi_1+\cos \theta_2 \dd \phi_2 \wedge \dd \psi_2, \quad e^{\Phi} = \frac{4}{R_+ R_-}, \nn\\
 F_3 &= \frac{R_+^2}{4} \sin \theta_1 \dd \theta_1 \wedge \dd \phi_1 \wedge \dd \psi_2-\frac{R_-^2}{4} \sin \theta_2 \dd \theta_2 \wedge \dd \phi_2 \wedge \dd \psi_1, \nn\\[2mm]
 F_5 &= (1+ *_{10}) \left[ - 2 L^2 \textrm{Vol}(AdS_3) \wedge \dd \psi_1 \wedge \dd \psi_2 \right],    
\end{align}
where we have added subscripts to distinguish the angular coordinates. We note that the NS sector is even under an exchange of angular coordinates, whereas the RR sector is odd. We now check the remaining supersymmetry. From the dilatino variation, we get the projection condition: 
\be
\label{proj1}
\Gamma^{\psi_1 \psi_2} i \sigma_2 \eta = - \eta. 
\ee
From the gravitino variations along the $x$, $\psi_1$ and $\psi_2$ directions, we get the additional projection
\bea
\label{proj2}
\left[ \frac{L}{R_+} \Gamma^{012 \theta_1 \phi_1 \psi_2}-\frac{L}{R_-} \Gamma^{012 \theta_2 \phi_2 \psi_1}  \right] \sigma_1 \eta &=& - \eta. 
\eea
One can check that the two projection conditions we have indeed commute, so supersymmetry is not broken further. 

We can once again solve for angular dependence, getting 
\be
\eta = e^{- \frac{1}{2} \theta_1 \Gamma^{\phi_1 \psi_1} \sigma_3} e^{-\frac{1}{2} \phi_1 \Gamma^{\phi_1 \theta_1} } e^{- \frac{1}{2} \theta_2 \Gamma^{\phi_2 \psi_2} \sigma_3} e^{-\frac{1}{2} \phi_2 \Gamma^{\phi_2 \theta_2} } \tilde{\eta}, 
\ee
where $\tilde{\eta}$ is expected to be the Killing spinor for $AdS_3$. Indeed, one can check that the remaining equation is just the Killing spinor equation for $AdS_3$, $\nabla_{\mu} \eta = \frac{1}{2} \gamma_3 \Gamma_{\mu} \eta$, where we have defined $\gamma_3 \equiv \Gamma^{012}$. Solving the $AdS_3$ Killing spinor equation, we find 
\be
\label{AdS_spinor}
\tilde{\eta} =  \left(r^{\frac{1}{2}} + r^{-\frac{1}{2}} (t \Gamma_{0}^{~2} + x_1 \Gamma_{1}^{~2} ) \right) \tilde{\eta}_+ + r^{-\frac{1}{2}} \tilde{\eta}_-, 
\ee
where $\tilde{\eta}_{\pm}$ are constant spinors subject to (\ref{proj1}) and (\ref{proj2}) satisfying $\Gamma^{01} \tilde{\eta}_{\pm} = \pm \tilde{\eta}_{\pm}$. We clearly see that the preserved supersymmetry is $\mathcal{N} = (0,4)$, since the Killing spinors separate into the usual Poincar\'e and superconformal Killing spinors, each with a different chirality.  The same conclusion can be drawn for the non-Abelian T-dual in section \ref{sec:SU(2)x2}. However, in contrast to the usual small superconformal symmetry, we appear to have $SU(2) \times SU(2)$ R-symmetry, which is suggested from the angular dependence of the Killing spinors. 

To further check the R-symmetry, we can also analyse the isometry algebra in our conventions in 10D, following a procedure outlined in ref. \cite{Gauntlett:1998kc}. The first step is to identify the corresponding generic 10D Killing vector field, whose existence is always guaranteed for supersymmetric geometries. Using the Killing spinor equations presented in section \ref{sec:SUSY}, standard arguments show that 
\be
V = \frac{1}{2} \left(  \bar{\epsilon}_{+} \Gamma^{M} \epsilon_+ + \bar{\epsilon}_- \Gamma^{M} \epsilon_{-} \right) \partial_{M} 
\ee
is always Killing. Note, we define $\bar{\epsilon} \equiv \epsilon^{\dagger} \Gamma^{0}$, with $(\Gamma^0)^{\dagger} = - \Gamma^0$ and $(\Gamma^{i})^{\dagger} = \Gamma^{i}, i = 1, \dots, 9$. From (\ref{proj1}), we have $\epsilon_- = \Gamma^{\psi_1 \psi_2} \epsilon_+$, and as a result, 
\be
V^{M} = \frac{1}{2} \left( \bar{\epsilon}_{+} \Gamma^{M} \epsilon_+ - \bar{\epsilon}_+ \Gamma^{\psi_1 \psi_2} \Gamma^{M} \Gamma^{\psi_1 \psi_2} \epsilon_{+} \right). 
\ee
We immediately recognise that $V^{\psi_i} = 0$, which is as expected, since these components of the vector field drop out when we reduce on a Hopf-fibre from 11D \cite{Gauntlett:1998kc}. Thus, $V^{M} = \bar{\epsilon}_{+} \Gamma^{M} \epsilon_+$, only depends on one of the Majorana-Weyl spinors. 

It is then a simple exercise to determine the internal components of $V$, 
\bea
V_{\rm int} &=& \frac{2}{R_+} \bar{\epsilon}_+ \Gamma^{\theta_1} \epsilon_- \, \xi^{+}_{1}+ \frac{2}{R_+} \bar{\epsilon}_+ \Gamma^{\phi_1} \epsilon_- \xi^{+}_2   
+ \frac{2}{R_-} \bar{\epsilon}_+ \Gamma^{\theta_2} \epsilon_- \xi^{-}_1+ \frac{2}{R_-} \bar{\epsilon}_+ \Gamma^{\phi_2} \epsilon_- \xi^{-}_2   \cr
&-& \frac{2}{R_+}  \bar{\epsilon}_+ \Gamma^{\psi_1} \epsilon_- \xi^{+}_3- \frac{2}{R_-} \bar{\epsilon}_+ \Gamma^{\psi_2} \epsilon_- \xi^{-}_3 + (+ \longleftrightarrow -),   
\eea
where we have relabeled $\tilde{\epsilon}$ simply $\epsilon$ for convenience and have defined the following two-sphere Killing vectors: 
\bea
\xi_{1}^{+} &=&   \cos \phi_1 \partial_{\theta_1} - \sin \phi_1 \cot \theta_1 \partial_{\phi_1}, ~~
\xi_{2}^{+} = \sin \phi_1 \partial_{\theta_1} + \cos \phi_1 \cot \theta_1 \partial_{\phi_1},  ~~
\xi_3^{+} = \partial_{\phi_1},   \nonumber 
\eea
with $\xi^{-}_i$ similarly defined in terms of the coordinates $(\theta_2, \phi_2)$. Note the Killing vectors satisfy the expected $SU(2)$ commutation relations, $[ \xi^{(+)}_i, \xi^{(+)}_j] = - \epsilon_{ijk}  \xi_k^{(+)}$, etc. 

The subscripts on $\epsilon$ refer to chirality with respect to $\Gamma^{01}$. It is straightforward to show that other combinations of spinors cannot contribute to these vector bilinears. We note that since we have eight supersymmetries, there is \textit{a priori} no relation between say $\bar{\epsilon}_+ \Gamma^{\theta_1} \epsilon_-$ and $\bar{\epsilon}_+ \Gamma^{\theta_2} \epsilon_-$ etc., so the $SU(2)$ symmetries should be viewed as being independent. This suggests an $\mathcal{N} = (0,4)$ SCFT with $SU(2) \times SU(2)$ R-symmetry.     

Evaluating the external $AdS_3$ components of the Killing vector $V$, we find 
\bea
V_{\rm ext} &=& (\bar{\epsilon}_+ \Gamma^2 \epsilon_- + \bar{\epsilon}_- \Gamma^2 \epsilon_+) \left( M_{01} +D \right) + \bar{\epsilon}_- \Gamma^0 \epsilon_- (P_0 - P_1) + \bar{\epsilon}_+ \Gamma^0 \epsilon_+ (K_0 + K_1), \nonumber
\eea
where now $\epsilon$ denotes the constant chiral spinors appearing in the expression for the $AdS_3$ Killing spinor (\ref{AdS_spinor}), and we have defined the $AdS_3$ Killing vectors in Poincar\'e patch as 
\bea
P_0 &=& \partial_{t}, \quad
P_1 = -\partial_{x}, \cr
M_{01} &=& x \, \partial_{t} + t \, \partial_{x}, \quad
D = r \, \partial_{r} + t \, \partial_{t} + x \, \partial_{x}, \cr
K_0 &=& (t^2 + x^2 + r^2) \partial_{t} + 2 t ( r \partial_{r} + x \partial_{x}), \cr 
K_1 &=& (t^2 + x^2 - r^2) \partial_{x} + 2 x ( r \partial_{r} + t \partial_{t}). 
\eea
These satisfy the usual conformal algebra:  
\bea
[M_{\mu \nu}, P_{\rho}] &=& - ( \eta_{\mu \rho} P_{\nu} - \eta_{\nu \rho} P_{\mu} ), ~~[M_{\mu \nu}, K_{\rho}] = - ( \eta_{\mu \rho} K_{\nu} - \eta_{\nu \rho} K_{\mu} ), \cr
[M_{\mu \nu}, D] &=& 0, ~~[D, P_{\mu}] = - P_{\mu}, ~~ [D, K_{\mu}] = K_{\mu}, ~~
[P_{\mu}, K_{\nu} ] = 2 M_{\mu \nu} - 2 \eta_{\mu \nu} D, 
\eea
with $\mu, \nu = 0, 1$. 

Recalling the bosonic subgroup of $D(2|1, \gamma)$, we come to the conclusion that after two Hopf-fibre T-dualities, the isometry supergroup of the $AdS_3 \times S^3 \times S^3$ geometry, namely $D(2|1, \gamma) \times D(2|1, \gamma)$ becomes simply $D(2|1, \gamma) \times SL(2, \mathbb{R})$, where $\gamma$ is the ratio of the two-sphere radii.

\end{document}